\documentclass[aps,superscriptaddress,preprintnumbers,showpacs,showkeys,nofootinbib,floatfix]{revtex4}
\usepackage{graphicx,epsfig,wrapfig,amssymb}
\newcommand{\be}{\begin{equation}}
\newcommand{\ee}{\end{equation}}
\newcommand{\ba}{\begin{eqnarray}}
\newcommand{\ea}{\end{eqnarray}}
\newcommand{\la}{\langle}
\newcommand{\ra}{\rangle}
\newcommand{\di}{ {\rm d} }
\begin{document}
\newcommand*{\Jlab}{Thomas Jefferson National Accelerator Facility,
Newport News, VA 23606, U.S.A.}\affiliation{\Jlab}
\newcommand*{\Dubna}{Joint Institute for Nuclear Research, Dubna,
141980 Russia}\affiliation{\Dubna}
\newcommand*{\Bochum}{Institut f{\"u}r Theoretische Physik II,
Ruhr-Universit{\"a}t Bochum, D-44780 Bochum, Germany}\affiliation{\Bochum}
\newcommand*{\BNL}{RIKEN BNL Research Center, Building 510A, BNL,
Upton, NY 11973, U.S.A.}\affiliation{\BNL}
\newcommand*{\Berkley}{Nuclear Science Division, Lawrence Berkeley National
Laboratory, Berkeley, CA 94720, U.S.A.}\affiliation{\Berkley}

\title{ \boldmath Pretzelosity distribution function $h_{1T}^\perp$
    and the single spin asymmetry $A_{UT}^{\sin(3\phi-\phi_S)}$}
\author{H.~Avakian}\affiliation{\Jlab}
\author{A.~V.~Efremov}\affiliation{\Dubna}
\author{P.~Schweitzer}\affiliation{\Bochum}
\author{F.~Yuan}\affiliation{\BNL}\affiliation{\Berkley}

\date{May 2008}
\begin{abstract}
  The leading twist transverse momentum dependent parton distribution function
  $h_{1T}^\perp$, which is sometimes called ``pretzelosity,'' is studied.
  We review the theoretical properties of this function, and present
  bag model predictions.
  We observe an interesting relation valid in a large class of relativistic models:
  The difference between helicity and transversity distributions,
  which is often said to be a 'measure of relativistic effects' in
  nucleon, is nothing but the pretzelosity distribution.
  Pretzelosity is chirally odd and can be accessed in combination with the
  Collins effect in semi-inclusive deep inelastic scattering, where it gives
  rise to an azimuthal single spin asymmetry proportional to $\sin(3\phi-\phi_S)$.
  We discuss the preliminary deuteron target data  from COMPASS,
  on that observable and make predictions for future experiments
  on various targets at JLab, COMPASS and HERMES.
 \end{abstract}
\maketitle
\section{Introduction}
\label{Sec-1:introduction}

Processes like semi-inclusive deep-inelastic lepton nucleon scattering (SIDIS),
hadron production in $e^+e^-$ 
annihilations or the Drell-Yan process
\cite{Cahn:1978se,Collins:1984kg,Sivers:1989cc,Efremov:1992pe,Collins:1992kk,Collins:1993kq,Kotzinian:1994dv,Mulders:1995dh,Boer:1997nt,Boer:1997mf,Boer:1997qn,Boer:1999mm,Brodsky:2002cx,Collins:2002kn,Belitsky:2002sm,Cherednikov:2007tw}
factorize at leading twist \cite{Collins:1981uk,Ji:2004wu,Collins:2004nx}
allowing to access info\-rmation on transverse momentum dependent fragmentation
and parton distribution functions (TMDs) \cite{Collins:2003fm,Collins:2007ph}.
The latter contain novel, so far unexplored information on the nucleon structure.
In order to be sensitive to ``intrinsic'' transverse parton momenta it is necessary
to measure adequate transverse momenta in the final state, for example, in SIDIS
the transverse momenta of produced hadrons with respect to the virtual photon.

The eight leading-twist TMDs $f_1^a$, $f_{1T}^{\perp a}$, $g_1^a$, $g_{1T}^a$,
$h_1^a$, $h_{1L}^{\perp a}$, $h_{1T}^{\perp a}$, $h_1^{\perp a}$ \cite{Boer:1997nt},
and further subleading-twist structures \cite{Goeke:2005hb,Bacchetta:2006tn}
describe the structure of the nucleon in these reactions. The fragmentation
of unpolarized hadrons is described in terms two fragmentation functions,
$D_1^a$ and $H_1^{\perp a}$, at leading-twist.
In SIDIS (with polarized beams and/or targets, where necessary) it is possible to
access information on the leading twist TMDs by measuring the angular distributions
of produced hadrons. Some data on such processes are available
\cite{Arneodo:1986cf,Airapetian:1999tv,Airapetian:2001eg,Airapetian:2002mf,Avakian:2003pk,Airapetian:2004tw,Alexakhin:2005iw,Diefenthaler:2005gx,Gregor:2005qv,Ageev:2006da,Avakian:2005ps,Airapetian:2005jc,Airapetian:2006rx,Abe:2005zx,Ogawa:2006bm,Martin:2007au,Diefenthaler:2007rj,Kotzinian:2007uv,Seidl:2008xc}.

The fragmentation functions and TMDs in SIDIS and other processes were
subject to numerous studies in literature
\cite{Kotzinian:1995cz,DeSanctis:2000fh,Anselmino:2000mb,Efremov:2001cz,Efremov:2001ia,Ma:2002ns,Bacchetta:2002tk,Yuan:2003wk,Efremov:2003eq,D'Alesio:2004up,Efremov:2004tp,Collins:2005ie,Collins:2005rq,Anselmino:2005nn,Vogelsang:2005cs,Efremov:2006qm,Anselmino:2007fs,Arnold:2008ap,Anselmino:2008sg,Gamberg:2007gb,Kotzinian:2006dw,Brodsky:2006hj,Burkardt:2007rv,Avakian:2007xa,Bacchetta:1999kz,Avakian:2007mv}.
This is true especially for the prominent transversity distribution $h_1^a$ or the
'naively time-reversal-odd' functions like the Sivers function $f_{1T}^{\perp a}$,
the Boer-Mulders function $h_1^{\perp a}$ and the Collins fragmentation function
$H_1^{\perp a}$. The so far probably least considered function is the 'pretzelosity'
distribution $h_{1T}^{\perp a}$.

The purpose of this note is therefore three-fold.
First, we will review what is known about $h_{1T}^{\perp a}$.
Second, we will calculate this function in the bag model.
Third, we will present estimates for the transverse target single spin asymmetry
proportional to $\sin(3\phi-\phi_S)$ in SIDIS in which $h_{1T}^{\perp a}$ enters,
and discuss the prospects to measure this asymmetry in experiments at Jefferson Lab.

\section{TMDs and SIDIS}
\label{Sec-2:TMDs-and-SIDIS}

Hard processes sensitive to parton transverse momenta like SIDIS
are described in terms of light-front correlators
\be\label{Eq:correlator}
    \phi(x,\vec{p}_T)_{ij} = \int\frac{\di z^-\di^2\vec{z}_T}{(2\pi)^3}\;e^{ipz}\;
    \la N(P,S)|\bar\psi_j(0)\,\{\mbox{gauge link}\}\,\psi_i(z)|N(P,S)\ra
    \biggl|_{z^+=0,\,p^+ = xP^+} \;.
    \ee
We use light-cone coordinates $a^\pm=(a^0\pm a^3)/\sqrt{2}$. In SIDIS the
singled our 3-direction is along the momentum of the hard virtual photon,
and transverse vectors like $\vec{p}_T$ are perpendicular to it.
The symbolically indicated gauge-link depends on the process
\cite{Collins:2002kn,Belitsky:2002sm,Cherednikov:2007tw}. In the nucleon rest
frame the polarization vector $S=(0,\vec{S}_T,S_L)$ with $\vec{S}_T^2+S_L^2=1$.

The information content of the correlator  (\ref{Eq:correlator}) is summarized by
eight leading-twist TMDs \cite{Boer:1997nt}, that can be projected out from the
correlator (\ref{Eq:correlator}) as follows
\ba
    \frac12\;{\rm tr}\biggl[\gamma^+ \;\phi(x,\vec{p}_T)\biggr]         &=&
    \hspace{5mm}f_1 -\frac{\varepsilon^{jk}p_T^j S_T^k}{M_N}\,f_{1T}^\perp
    \label{Eq:TMD-pdfs-I}\\
    \frac12\;{\rm tr}\biggl[\gamma^+\gamma_5 \;\phi(x,\vec{p}_T)\biggr] &=&
    S_L\,g_1 + \frac{\vec{p}_T\cdot\vec{S}_T}{M_N}\,g_{1T}^\perp
    \label{Eq:TMD-pdfs-II}\\
    \frac12\;{\rm tr}\biggl[i\sigma^{j+}\gamma_5 \;\phi(x,\vec{p}_T)\biggr] &=&
    S_T^j\,h_1  + S_L\,\frac{p_T^j}{M_N}\,h_{1L}^\perp +
    \frac{(p_T^j p_T^k-\frac12\,\vec{p}_T^{\:2}\delta^{jk})S_T^k}{M_N^2}\,
    h_{1T}^\perp + \frac{\varepsilon^{jk}p_T^k}{M_N}\,h_1^\perp \;,
    \label{Eq:TMD-pdfs-III}
\ea
where the space-indices $j,k$ refer to the plane transverse with respect to the
light-cone and $\varepsilon^{12} = - \varepsilon^{21} = 1$ and zero else.
Integrating out transverse momenta in the correlator (\ref{Eq:correlator})
leads to the three 'usual' parton distributions known from collinear kinematics
$j_1^a(x) = \int\di^2\vec{p}_T \, j_1^a(x,\vec{p}_T^{\:2})$ with $j=f,\,g,\,h$
\cite{Ralston:1979ys,Jaffe:1991ra}. Dirac-structures other than that in
Eqs.~(\ref{Eq:TMD-pdfs-I},~\ref{Eq:TMD-pdfs-II},~\ref{Eq:TMD-pdfs-III})
lead to subleading-twist terms \cite{Goeke:2005hb,Bacchetta:2006tn}.
The fragmentation of unpolarized hadrons is described by a correlator
analog to (\ref{Eq:correlator}) parameterized in terms of two leading-twist
fragmentation functions, $D_1^a$ and $H_1^{\perp a}$.

\begin{figure}[b!]
    \vspace{1.1cm}
    \centering
        \includegraphics[width=9cm]{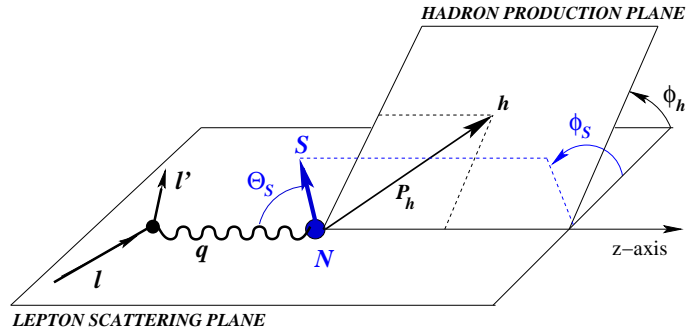}
        \caption{\label{fig1-processes-kinematics}
    Kinematics of the SIDIS process $lN\to l^\prime h X$
    and the definitions of azimuthal angles in the lab frame.}
\end{figure}

The process of SIDIS is sketched in Fig.~\ref{fig1-processes-kinematics}.
The cross section differential in the azimuthal angle $\phi_h$ of the
produced hadron, see Fig.~\ref{fig1-processes-kinematics},
(and possibly differential also in other kinematic variables)
has the following general decomposition \cite{Kotzinian:1994dv,Diehl:2005pc},
where the dots indicate power suppressed terms,
\ba\label{Eq:sigma-in-SIDIS}
    \frac{\di\sigma}{\di\phi_h}
    &=&                       F_{UU}
    +        \cos(2\phi_h)\,  F_{UU}^{\cos(2\phi_h)}
    + S_L    \sin(2\phi_h)\,  F_{UL}^{\sin(2\phi_h)}
    + \lambda\;\biggl[\;S_L   F_{LL}
    + S_T\cos(\phi_h-\phi_S)\,F_{LT}^{\cos( \phi_h-\phi_S)}\,\biggr] \nonumber\\
    &&\hspace{0.75cm}
    + \;S_T\biggl[\;
      \sin( \phi_h-\phi_S)\,  F_{UT}^{\sin( \phi_h-\phi_S)}
    + \sin( \phi_h+\phi_S)\,  F_{UT}^{\sin( \phi_h+\phi_S)}
    + \sin(3\phi_h-\phi_S)\,  F_{UT}^{\sin(3\phi_h-\phi_S)} \biggr]
    + \dots \;\;\; \label{Eq:azim-distr-in-SIDIS}
\ea
In $F_{XY}^{\rm weight}$ the first index $X=U(L)$ denotes the unpolarized beam
(longitudinally polarized beam with helicity $\lambda$).
The second index $Y=U(L,T)$ denotes the unpolarized target (longitudinally,
transversely with respect to the virtual photon polarized target).
The superscript reminds on the kind of angular distribution of the produced
hadrons with no index indicating an isotrop $\phi$-distribution, and
$\phi_S$ is the azimuthal angle of the transversely polarized target.

Each structure function arises from a different TMD.
The chirally even $f$'s and $g$'s enter the observables in connection with
the unpolarized fragmentation function $D_1^a$, the chirally odd $h$'s in
connection with the chirally odd Collins fragmentation function~$H_1^{\perp a}$.
One has
\ba
\label{Eq:FUU}
F_{UU}                   \propto\sum_ae_a^2\;f_{1 }^{      a}\otimes\,D_1^{      a}&,&
F_{LT}^{\cos( \phi_h-\phi_S)}\propto\sum_ae_a^2\;g_{1T}^{\perp a}\otimes\,D_1^{      a}\\
F_{LL}               \propto\sum_ae_a^2\;g_{1 }^{      a}\otimes\,D_1^{      a}&,&
F_{UT}^{\sin( \phi_h-\phi_S)}\propto\sum_ae_a^2\;f_{1T}^{\perp a}\otimes\,D_1^{      a}\\
F_{UU}^{\cos(2\phi_h)}       \propto\sum_ae_a^2\;h_{1 }^{\perp a}\otimes\,H_1^{\perp a}&,&
F_{UT}^{\sin( \phi_h+\phi_S)}\propto\sum_ae_a^2\;h_{1 }^{      a}\otimes\,H_1^{\perp a}\\
F_{UL}^{\sin(2\phi_h)}       \propto\sum_ae_a^2\;h_{1L}^{\perp a}\otimes\,H_1^{\perp a}&,&
F_{UT}^{\sin(3\phi_h-\phi_S)}\propto\sum_ae_a^2\;h_{1T}^{\perp a}\otimes\,H_1^{\perp a}
\label{AUTsin(3phi-phiS)}
\ea
More precisely, TMDs and fragmentation functions enter the respective
{\sl tree-level} expressions in certain convolution integrals,
indicated by $\otimes$ in (\ref{Eq:FUU}-\ref{AUTsin(3phi-phiS)}), which
entangle transverse parton momenta from TMDs and fragmentation functions.
(Going beyond tree-level description requires introduction of
soft factors \cite{Ji:2004wu,Collins:2004nx} from which we refrain here.)

In general, such convolution integrals cannot be solved, unless one weights
the DIS counts with an adequate power of transverse hadron momentum.
For the $\sin(3\phi_h-\phi_S)$ asymmetry that would mean to weight the
events with $P_{h\perp}^3$. The resulting model-independent tree-level
expression for that structure function is then given in terms of certain
transverse moments of $h_{1T}^\perp$ and $H_1^\perp$ \cite{Boer:1997nt}.

The analysis of such single spin asymmetries (SSA), however, is involved
because acceptance effects are difficult to control. So far only preliminary
data on (some different) $P_{h\perp}$-weighted SSAs are available \cite{Gregor:2005qv}.
Moreover, the $P_{h\perp}$-weighting obscures the actual size of the effect.
We shall therefore consider here the 'unweighted' SSA
\be\label{Eq:SSA-unweighted-I}
    A_{UT}^{\sin(3\phi-\phi_S)}
    = \frac{\overline{F}_{UT}^{\sin(3\phi-\phi_S)}}{\overline{F}_{UU}}\;,
    \ee
dependence with the known $y$-dependence removed, i.e.\
$F_{UT}^{\sin(3\phi-\phi_S)} = (1-y)\overline{F}_{UT}^{\sin(3\phi-\phi_S)}$
and  $F_{UU} = (1-y+y^2/2)\overline{F}_{UU}$, such that the SSA
depends only on $x$ and $z$.
(The SSA depends, of course, also on $Q^2$. In the parton model this dependence is
weak, logarithmic, and 'hidden' in the scale dependence of the distribution and
fragmentation functions, which we do not indicate for brevity.)

In this case, however, the integrals convoluting the transverse momenta can only
be solved, if one assumes some model for the transverse momentum dependence.
Here we shall assume the Gaussian Ansatz
\ba\label{Eq:Gauss-ansatz}
    h_{1T}^{\perp a}(x,\vec{p}_T^{\:2}) =
    h_{1T}^{\perp a}(x)\;
    \frac{\exp(-\vec{p}_T^{\:2}/p_{\rm av}^2)}{\pi\;p_{\rm av}^2} \;,\;\;\;
    H_1^{\perp a}(z,\vec{K}_T^{\:2}) =
    H_1^{\perp a}(z)\;
    \frac{\exp(-\vec{K}_T^{\:2}/K^2_{\rm av})}{\pi\;K^2_{\rm av}}\;.
    \ea
This is, of course, a crude approximation. However, besides being convenient
\cite{Mulders:1995dh}, this Ansatz is also phenomenologically useful, provided
the transverse hadron momenta are small compared to the relevant hard scale,
$\la P_{h\perp}\ra\ll Q$ in SIDIS, and one is interested in catching the gross
features of the effects \cite{D'Alesio:2004up}.
A high precision description of $p_T$-effects requires methods along the
QCD-based formalism of \cite{Collins:1984kg}, see \cite{Landry:2002ix} and
references therein for examples.

With the Ansatz (\ref{Eq:Gauss-ansatz}) we obtain for the SSA the following result
\be\label{Eq:AUT-Gauss}
    A_{UT}^{\sin(3\phi-\phi_S)}(x,z)
    = \frac{C_{\rm Gauss}\sum_a e_a^2 x \,h_{1T}^{\perp (1)a}(x)\,
      H_1^{\perp (1/2)a}(z)} {\sum_ae^a\,xf_1(x)\,D_1^a(z)}
    \;, \ee
which we have written such that the dependence on the Gauss model is 'dumped'
into the factor
\be\label{Eq:AUT-Gauss-factor}
    C_{\rm Gauss} = f(zp_{\rm av}/K_{\rm av}) \;\;\;\mbox{with}\;\;\;
    f(a) = \frac{3}{(a^{2/3}+a^{-2/3})^{3/2}} \;.
\ee
In (\ref{Eq:AUT-Gauss}) we have introduced the transverse moments
\be\label{Eq:trans-mom-1}
    h_{1T}^{\perp (1)a}(x) = \int\di^2\vec{p}_T\;\frac{\vec{p}_T^{\:2}}{2M_N^2}\;
    h_{1T}^{\perp    a}(x,\vec{p}_T^{\:2})\;,\;\;\;
    H_1^{ \perp (1/2)a}(z) = \int\di^2\vec{K}_T\;\frac{|\vec{K}_T|}{2zm_h}\;
    H_1^{ \perp      a}(z,\vec{K}_T^{\:2})\;.
\ee
Little is known about the  Gauss model parameters, $p_{\rm av}$ and $K_{\rm av}$.
Apart from constraints from positivity, we know nothing about
$p_{\rm av}$ and have only vague constraints on the Gaussian width of the
Collins function \cite{Efremov:2006qm,Anselmino:2007fs}. It is therefore
gratifying to observe that $f(a)$ in (\ref{Eq:AUT-Gauss-factor}) is a slowly
varying function of $a$ satisfying $0\le f(a)\le 3/(2\sqrt{2})$.
This means that
\be\label{Eq:AUT-Gauss-factor-max}
    0\le C_{\rm Gauss} \le C_{\rm max} = \frac{3}{2\sqrt{2}} \;.
\ee
For physically reasonable values of the ratio $p_{\rm av}/K_{\rm av}$
(let us guess, for example, that it coincides within a factor of two
with the corresponding ratio of the unpolarized widths
$p_{\rm unp}/K_{\rm unp}\approx 1.4$ \cite{Collins:2005ie}) one may
therefore expect that the factor $C_{\rm Gauss}$ is within 20$\,\%$
close to its maximum value.
In the following, when we will be interested in estimating the maximal
effects for the SSA (\ref{Eq:AUT-Gauss}), we will approximate
$C_{\rm Gauss}$ by  $C_{\rm max}$ in Eq.~(\ref{Eq:AUT-Gauss-factor-max}).

\newpage
\section{\boldmath What do we know about $h_{1T}^{\perp a}\,$?}
\label{Sec-3:What-we-know}

Let us summarize briefly, what we already know, or will learn in the
next sections, about the pretzelosity distribution.
\begin{enumerate}
\renewcommand{\labelenumi}{\sl (\arabic{enumi})$\,$}
\item   It is chirally odd.
\item   It has a probabilistic interpretation \cite{Mulders:1995dh}.
\item   At large-$x$ it is predicted to be suppressed by $(1-x)^2$ compared to
    $f_1^a(x)$ \cite{Brodsky:2006hj,Burkardt:2007rv,Avakian:2007xa}.
\item\label{point-small-x}
    It is expected to be suppressed also at small $x$ compared to $f_1^a$.
\item\label{point-positivity}
    It must satisfy the positivity condition \cite{Bacchetta:1999kz}
    \be\label{Eq:positivity}
    \biggl|\frac{\vec{p}_T^{\:2}}{2M_N^2}h_{1T}^{\perp a}(x,\vec{p}_T^{\:2})\biggr|
    \le \frac12\biggl(f_1^a(x,\vec{p}_T^{\:2})-g_1^a(x,\vec{p}_T^{\:2})\biggr)
    \le f_1^a(x,\vec{p}_T^{\:2})
    \ee
    or, after integrating over transverse momenta:
    \be\label{Eq:positivity-integrated}
    | h_{1T}^{\perp(1)a}(x)|
    \le \frac12\biggl(f_1^a(x) - g_1^a(x)\biggr) \le f_1^a(x)\;.
    \ee
\item\label{point-positivity-II}
    Adding up (\ref{Eq:positivity-integrated}) and the Soffer inequality
    $|h_1^a(x)|\le\frac12(f_1^a + g_1^a)(x)$ \cite{Soffer:1994ww},
    we obtain the remarkable bound:
    \be\label{Eq:positivity-with-h1}
    | h_{1T}^{\perp(1)a}(x)|+ |h_1^a(x)| \le f_1^a(x)\;.
    \ee
\item\label{point-large-Nc}
    In the limit of a large number of colors $N_c$ in QCD pretzelosity is predicted,
    for $p_T\sim {\cal O}(N_c^0)$ and $xN_c \sim {\cal O}(N_c^0)$, to exhibit
    the following flavor dependence (the same relation holds for antiquarks)
    \cite{Pobylitsa:2003ty}
    \be\label{Eq:large-Nc}
    \underbrace{|h_1^{\perp(1)u}(x)-h_1^{\perp(1)d}(x)|}_{{\cal O}(N_c^2)} \gg
    \underbrace{|h_1^{\perp(1)u}(x)+h_1^{\perp(1)d}(x)|}_{{\cal O}(N_c)}
    \ee
\item   There are estimates from spectator model \cite{Jakob:1997wg},
    and bag model, see Sec.~\ref{Sec-4:pretzelosity-in-bag}.
\item   In a large class of models, including the bag and spectator models,
    pretzelosity is the difference of helicity and transversity
    distributions, and in this sense a measure for relativistic effects,
    see Secs.~\ref{Sec-4:pretzelosity-in-bag} and \ref{Sec-5:in-spectator-model}.
\item   In some sense it 'measures' the deviation of the 'nucleon shape'
    from a sphere \cite{Miller:2007ae}.
\item   In simple (spectator-type) models, it has been related to chirally
    odd generalized parton distributions \cite{Meissner:2007rx}.
\item   It requires the presence of nucleon wave-function components with
    two units orbital momentum difference, e.g.\  $s$-$d$ interference
    of quadratic in $p$-wave component \cite{Burkardt:2007rv}.
\item\label{point-h1Tperp-gluon}
    There is no gluon analog of pretzelosity.
\end{enumerate}
Some comments are in order.

Concerning the point (\ref{point-small-x}):
This expectation is based on experience with evolution properties of integrated
parton distributions. The evolution of the chiral-odd $h_1^a$ differs significantly
from that of the chiral-even $f_1^a$ or $g_1^a$. If at some initial scale $h_1^a$
were as large as $f_1^a$ or $g_1^a$ (or the Soffer bound \cite{Soffer:1994ww})
in the small-$x$ region, then at higher scales it would be suppressed
\cite{Artru:1989zv,h1-evolve-NLO}. This pattern is expected also
for other chirally odd TMDs.

Concerning the point (\ref{point-positivity-II}):
For $u$-quarks, that usually dominate in the proton in the valence-$x$ region,
the bound (\ref{Eq:positivity-integrated}) is more restrictive than the Soffer
bound for transversity, because $g_1^u(x)$ is positive. Thus, positivity
gives more ``room'' for $|h_1^u(x)|$ compared to $| h_{1T}^{\perp(1)u}(x)|$
(and vice versa for the $d$-flavor).
But from this observation alone we cannot draw any conclusion on the
relative size of $|h_1^a(x)|$ and $| h_{1T}^{\perp(1)a}(x)|$, because
neither function is a priori ``forced'' to saturate its allowed
positivity bound. However, there are first indications that $h_1^u(x)$
is large, definitely larger than $\frac12(f_1^u - g_1^u)(x)$
\cite{Vogelsang:2005cs,Efremov:2006qm,Anselmino:2007fs}.
So one may expect $| h_{1T}^{\perp(1)u}(x)|\lesssim h_1^u(x)$ on the basis
of the first indications \cite{Vogelsang:2005cs,Efremov:2006qm,Anselmino:2007fs}.

Concerning the point (\ref{point-large-Nc}):
The same flavor dependence as in Eq.~(\ref{Eq:large-Nc}) holds for all
TMDs in a polarized nucleon. Notice, that there is also an unintegrated
version of (\ref{Eq:large-Nc}), see \cite{Pobylitsa:2003ty}.

Concerning the point (\ref{point-h1Tperp-gluon}):
Actually, in the decomposition of the gluon analog of the correlator
(\ref{Eq:correlator}) a structure appears that in Ref.~\cite{Meissner:2007rx}
has been called $h_{1T}^{\perp g}$, which was of notational convenience for
that work. (In Ref.~\cite{Mulders:2000sh} it was given a different name.)
This gluon TMD, however, has different properties compared to our quark
pretzelosity. For example, the $h_{1T}^{\perp g}$ of \cite{Meissner:2007rx}
is 'odd' under time-reversal while $h_{1T}^{\perp a}$ with $a=q,\;\bar q$
is 'even'.

\newpage
\section{\boldmath Pretzelosity in the bag model}
\label{Sec-4:pretzelosity-in-bag}

In the MIT bag model, the quark field has the following general form
\cite{Chodos:1974je,Jaffe:1974nj,Celenza:1982uk},
\begin{equation}\label{bw}
    \Psi_\alpha(\vec{x},t)=\sum\limits_{n>0,\kappa=\pm 1,m=\pm 1/2} N(n\kappa)
    \{ b_\alpha(n\kappa m)\psi_{n\kappa jm}(\vec{x},t)+
    d_\alpha^\dagger(n\kappa m)\psi_{-n-\kappa jm}(\vec{x},t)\} \ ,
    \end{equation}
where $b_\alpha^\dagger$ and $d_\alpha^\dagger$ create quark and anti-quark
excitations in the bag with the wave functions
\begin{equation}
    \psi_{n,-1,\frac{1}{2}m}(\vec{x},t)=\frac{1}{\sqrt{4\pi}}
    \left ( \begin{array}{r}
    i j_0(\frac{\omega_{n,-1}|\vec{x}|}{R_0})\chi_m\\
    -\vec{\sigma}\cdot\hat{{x}} \; j_1(\frac{\omega_{n,-1}|\vec{x}|}{R_0})\chi_m
    \end{array} \right )
    e^{-i\omega_{n,-1} t/R_0} \ .
    \end{equation}
For the lowest mode, we have $n=1$, $\kappa=-1$, and $\omega_{1,-1}\approx 2.04$
denoted as $\omega\equiv\omega_{1,-1}$ in the following. In the above equation,
$\vec{\sigma}$ is the $2\times 2$ Pauli matrix, $\chi_m$ the Pauli spinor, $R_0$
the bag radius, $\hat{{x}}=\vec{x}/|\vec{x}|$, and $j_i$, are spherical Bessel
functions. Taking the Fourier transformation, we have the momentum space
wave function for the lowest mode,
\begin{equation}
    \varphi_{m}(\vec{k})=i\sqrt{4\pi}N R_0^3
    \left (\begin{array}{r} t_0(k)\chi_m\\
    \vec{\sigma}\cdot\hat{k} \;t_1(k)\chi_m
    \end{array} \right ) \ ,
    \label{wp}
    \end{equation}
where $\hat{k} = \vec{k}/k$ with $k=|\vec{k}|$ and the normalization factor $N$ is,
\begin{equation}
        N=\left(\frac{\omega^3}{2R_0^3(\omega-1)\sin^2\omega}\right)^{1/2} \ .
    \end{equation}
The two functions $t_i$, $i=0,1$ are defined as
\begin{equation}
    \label{Eq:t0-t1}
    t_i(k)=\int\limits_0^1 u^2 du j_i(ukR_0)j_i(u\omega) \ .
\end{equation}
From the above equations, we see that the bag model wave function Eq.~(\ref{wp})
contains both $S$ and $P$ wave components. Especially, $t_0$ represents the $S$-wave
component, whereas $t_1$ represents the $P$-wave component of the proton wave
functions.

With the above wave functions, we can calculate all the leading order
TMD quark distributions. Assuming $SU(6)$ spin-flavor symmetry of the
proton wave function, we have the up-quark and down-quark distributions,
\begin{equation}\label{Eq:wafe-function-SU(6)-pol}
    h_{1T}^{\perp q}(x,k_\perp)= P_q \, h_{1T}^\perp(x,k_\perp),~~~
    P_u =  \frac{4}{3}\;,\;\;\;
    P_d = -\frac{1}{3}\;,
\end{equation}
where $h_{1T}^\perp$ can be written as,
\be
    h_{1T}^\perp(x,k_\perp) = A\,\biggl[-\frac{2M_N^2 t_1^2}{k^2}\biggr]
    \ , \;\;\; A = \frac{16\omega^4}{\pi^2(\omega -1)j_0^2(\omega)\,M_N^2}
\ee
where $\omega$ is the lowest root of the bag eigen-equation as we mentioned
above. The momenta $k_z$ and $k$ are defined as $k_z=xM_N-\omega/R_0$, and
$k=\sqrt{k_z^2+k_\perp^2}$. $R_0$ and $M_N$ are bag radius and proton mass,
respectively. In our calculations, we fix the dimensionless parameter
$R_0 M_N=4\omega$. Because $h_{1T}^\perp$ depends on $t_1^2$, clearly
it is proportional to the squared of the $P$-wave~\cite{Burkardt:2007rv},
and is sensitive to the quark orbital angular momentum in the proton.

In order to study the positivity bounds we have discussed in the last section, we
list the unpolarized and polarized quark distributions in the MIT bag model.
We have
\ba
   f_1(x,k_\perp) &=& A\biggl[t_0^2+2t_0t_1\frac{k_z}{k}+t_1^2 \biggr] \\
   g_1(x,k_\perp) &=& A\biggl[t_0^2+2t_0t_1\frac{k_z}{k}+t_1^2
            (\frac{2k_z^2}{k^2}-1) \biggr]  \\
   h_1(x,k_\perp) &=& A\biggl[t_0^2+2t_0t_1\frac{k_z}{k}+t_1^2\frac{k_z^2}{k^2}\biggr]
\ea
where according to the SU(6) spin flavor symmetry
\be\label{Eq:wafe-function-SU(6)-unp}
    f_1^q(x,k_\perp)=N_q f_1(x,k_\perp)\, , \;\;\;
    N_u = 2\, , \;\;\;
    N_d = 1\, ,
\ee
and the polarized TMDs of definite flavor, $g_1^q$ and $h_1^q$, are given by the
expressions analog to (\ref{Eq:wafe-function-SU(6)-pol}).

From the above results, we read off the relations, for example
\ba
&&  f_1(x,k_\perp) + g_1(x,k_\perp) = \;2 h_1(x,k_\perp)
    \label{Eq:specific-rel-I}\\
&&  h_1(x,k_\perp) - h_{1T}^{\perp(1)}(x,k_\perp) = f_1(x,k_\perp)
    \label{Eq:specific-rel-II}\ea
where $h_{1T}^{\perp(1)}(x,k_\perp) = k_\perp^2/(2M_N^2)\,h_{1T}^{\perp}(x,k_\perp)$.
Thus, in the bag model out of the four functions $f_1$, $g_1$, $h_1$, $h_{1T}^\perp$
only two are linearly independent.

In QCD the various TMDs are all independent of each other, and describe different
aspects of the nucleon structure. In simple models, however,
it is natural to encounter relations among TMDs. Since they are all
expressed in terms of $t_0$ and~$t_1$ representing the S and P-wave components
of the proton wave function, Eqs.~(\ref{wp},~\ref{Eq:t0-t1}),
the TMDs cannot be all linearly independent of each other. The specific
form of the relations (\ref{Eq:specific-rel-I},~\ref{Eq:specific-rel-II})
can be traced back to Melosh rotations, which relate longitudinal and transverse
nucleon polarization states in a Lorentz-invariant way \cite{Efremov:2002qh}.
The relations (\ref{Eq:specific-rel-I},~\ref{Eq:specific-rel-II})
arise therefore in some sense from Lorentz invariance properly took into
account in a simplified description of the nucleon.

Considering the flavor factors
(\ref{Eq:wafe-function-SU(6)-pol},~\ref{Eq:wafe-function-SU(6)-unp}) dictated by
SU(6) symmetry, we obtain from (\ref{Eq:specific-rel-I},~\ref{Eq:specific-rel-II})
the relations
\ba
&&  \frac{P_q}{N_q}\,f_1^q(x,k_\perp) + g_1^q(x,k_\perp) = \;2 h_1^q(x,k_\perp)
    \label{Eq:specific-rel-I-flavour} \\
&&   h_1^q(x,k_\perp)-h_{1T}^{\perp(1)q}(x,k_\perp)
    = \frac{P_q}{N_q}\,f_1^q(x,k_\perp) \label{Eq:specific-rel-II-flavour}\;.
\ea
The latter can be rewritten as follows
(notice that the signs are such that the first part really remains an equality)
\be
    |h_1^q(x,k_\perp)|+|h_{1T}^{\perp(1)q}(x,k_\perp)|
    = \frac{|P_q|}{N_q} f_1^q(x,k_\perp) < f_1^q(x,k_\perp)
    \label{Eq:specific-ineq}\;.
\ee
Comparing (\ref{Eq:specific-ineq}) with (\ref{Eq:positivity-with-h1})
we see that the inequality (\ref{Eq:positivity-with-h1}) is always satisfied
in the bag model, and actually never becomes an equality. In fact, transversity
and pretzelosity explore only $\frac23$ (only $\frac13$) of their allowed
positivity bound for the $u$-flavor ($d$-flavor).
Also the inequality (\ref{Eq:positivity}) is satisfied, because the Soffer
inequality holds in the bag model, which follows analogously from
(\ref{Eq:specific-rel-I-flavour}), cf.\  also the detailed discussion in
\cite{Barone:2001sp}.

Concerning large-$N_c$, we recall that in the quark model
formulated for a general (odd) number of colors $N_c$, the flavor factors
(\ref{Eq:wafe-function-SU(6)-pol},~\ref{Eq:wafe-function-SU(6)-unp})
are given by $P_u = (N_c+5)/6$ and $P_d=(-N_c+1)/6$ while
$N_u = (N_c+1)/2$ and $N_d=(N_c-1)/2$ \cite{Karl:1984cz}.
Thus, the bag model results respect large-$N_c$ predictions
(\ref{Eq:large-Nc}) because the employed SU(6) symmetry does so.
We also see, however, that $1/N_c$-corrections are sizeable for $N_c=3$.

Since only two out of the 4 functions $f_1$, $g_1$, $h_1$, $h_{1T}^\perp$
are linearly independent there are numerous relations among them.
The probably most interesting relation results from subtracting
the relations (\ref{Eq:specific-rel-I},~\ref{Eq:specific-rel-II})
from each other. This yields (since here a common flavor factor,
$P_q$, enters) the result
\be\label{Eq:measure-of-relativity}
    g_1^q(x,k_\perp) - h_1^q(x,k_\perp)
    = h_{1T}^{\perp(1)q}(x,k_\perp)\;.
\ee
This is result is remarkable from the point of view of the popular
statement 'the difference between transversity and helicity distributions
is a measure for relativistic effects in nucleon' \cite{Jaffe:1991ra}.
Here, in the framework of the bag model, we can quantify this statement:
The difference of helicity and transversity is (a transverse moment of)
pretzelosity.

From this discussion we immediately see what happens to $h_{1T}^{\perp q}$
in the non-relativistic limit: it vanishes. In this sense, we arrive at the
conclusion that the pretzelosity distribution itself is a measure for
relativistic effects in nucleon.

\begin{figure}[b]
        \includegraphics[width=7.5cm]{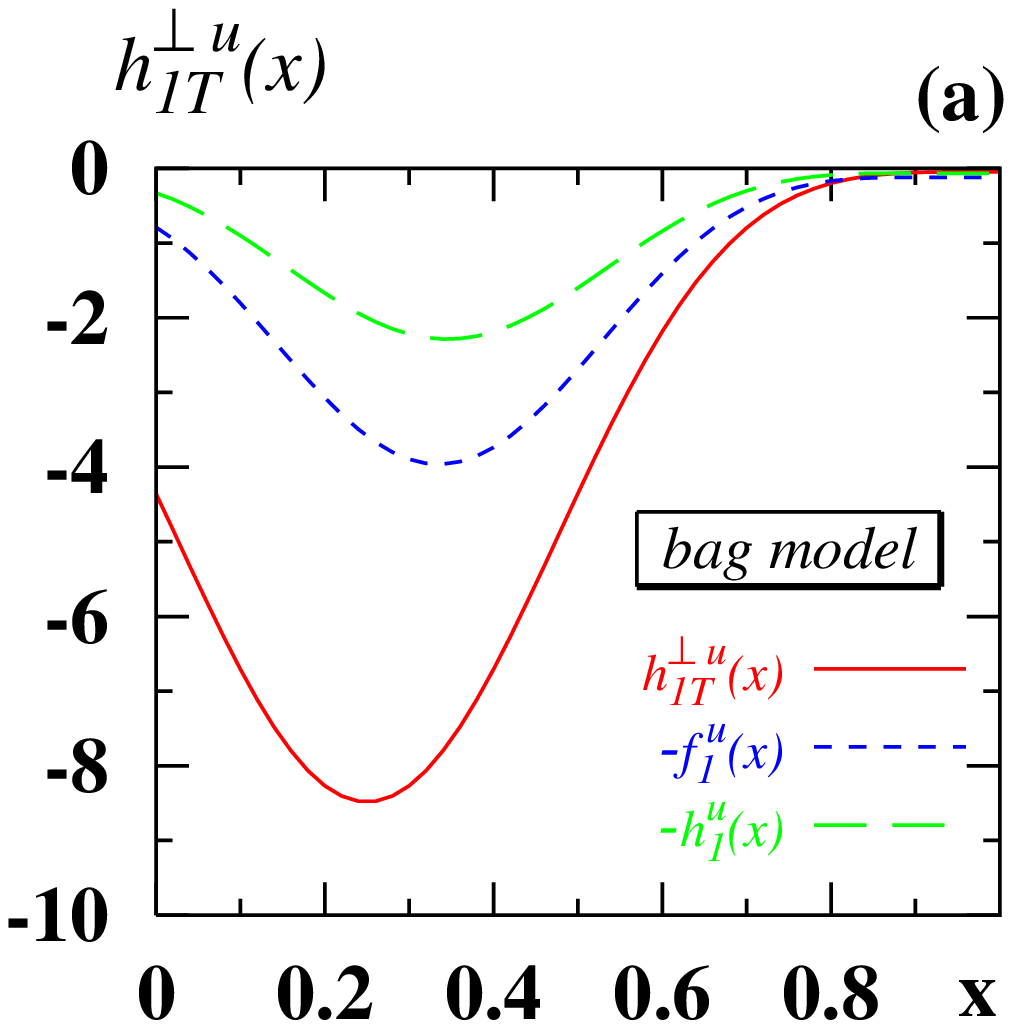}
        \includegraphics[width=7.5cm]{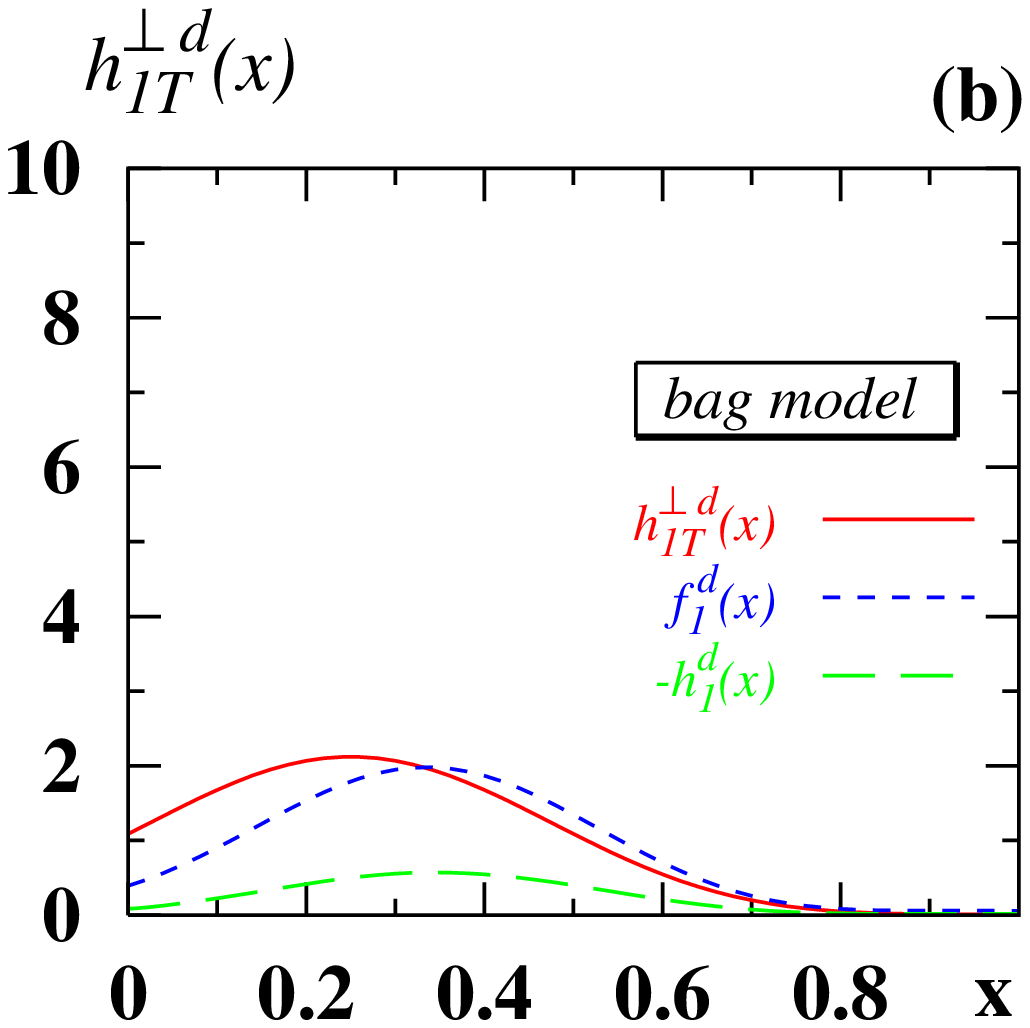}
\caption{\label{Fig02:h1Tperp-bag-x}
    The parton distribution function $h_{1T}^{\perp q}(x)$ vs.\ $x$
    from the bag model (results obtained here) in comparison
    to $f_1^q(x)$ and $h_1^q(x)$ from the same model.
    The functions $h_{1T}^{\perp q}(x)$ are rather large, even larger
    than $f_1^q(x)$. Notice, however, that $h_{1T}^{\perp q}(x)$ itself,
    as defined in (\ref{Eq:def-h1Tperp-of-x}), is not constrained
    by positivity bounds. All results refer to the low scale of the
    bag model.}
        \includegraphics[width=7.5cm]{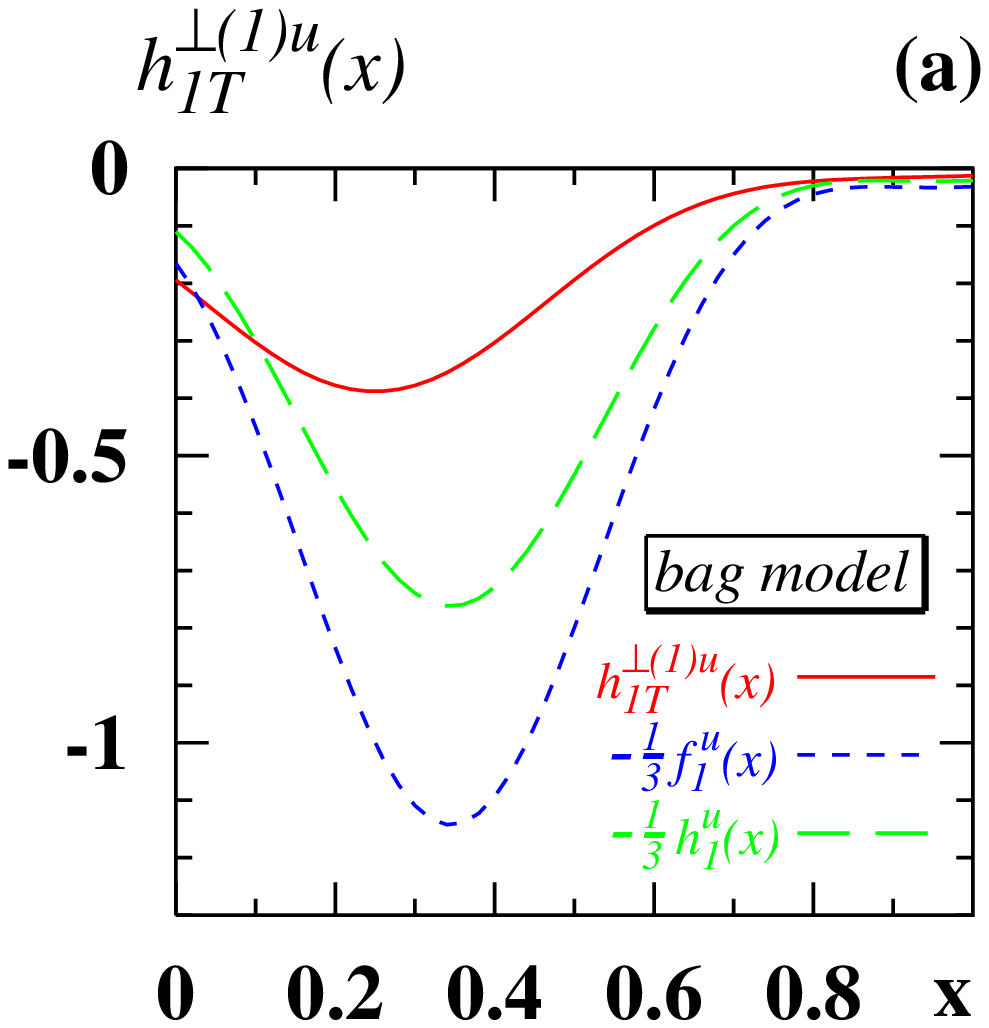}
        \includegraphics[width=7.5cm]{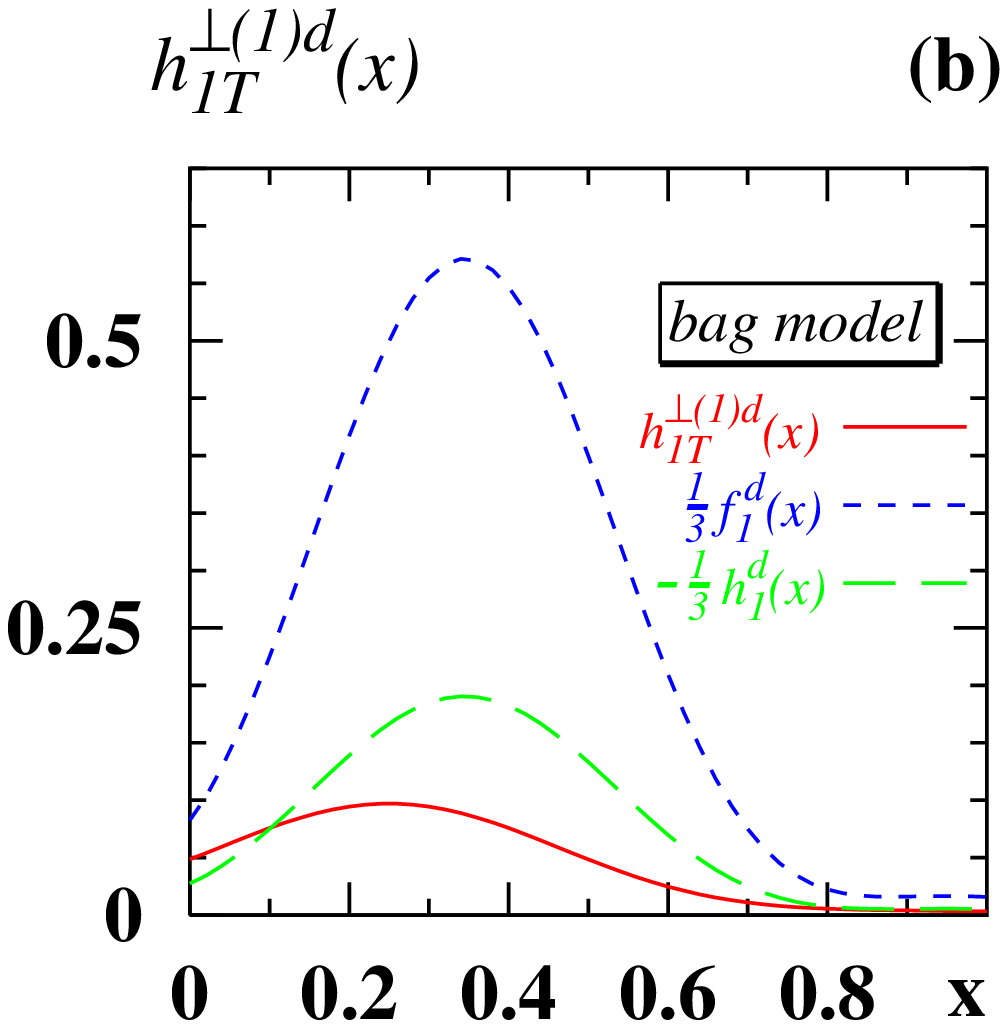}
\caption{\label{Fig03:h1Tperp1-bag-x}
    The transverse moment $h_{1T}^{\perp(1)q}(x)$ vs.\ $x$
    from the bag model (results obtained here) in comparison
    to $f_1^q(x)$ and $h_1^q(x)$ from the same model
    (rescaled by the factor $\pm\frac13$ for better visibility).
    The functions $h_{1T}^{\perp(1)q}(x)$ satisfy the positivity bounds,
    Eqs.~(\ref{Eq:positivity-integrated},~\ref{Eq:positivity-with-h1}).
    The results refer to the low scale of the bag model.}
\end{figure}

Let us turn to the discussion of the numerical results\footnote{
    \label{footnote:qbar-in-bag-model}
    The MIT bag gives also rise to antiquark distributions, though
    to unphysical ones as the $f_1^{\bar q}(x)$ come out negative,
    violating positivity. Nevertheless these contributions are crucial
    for the normalizations $\int\di x\,f_1^q(x) = N_q$ or the momentum
    sum rule $\sum_q\int\di x\,xf_1^q(x) = 1$, that are satisfied only
    if one includes the contribution from negative $x$
    (where $f_1^q(x)$ means minus the antiquark-distribution).
    Moreover, the TMDs receive non-vanishing support also from
    $|x|>1$, and the sum rules are satisfied only when integrating
    over the whole $x$-axis.
    It has been discussed in literature how to deal with these caveats,
    see for example \cite{Schreiber:1991tc}.
    In this work, we limit ourselves to quark TMDs at $0\le x\le 1$
    (we do not discuss 'valence distributions'
    $f_{1\,\rm val}^q(x)=f_1^q(x)-f_1^{\bar q}(x)$).
    When discussing sum rules, however, we shall integrate the TMDs
    over the whole $x$-axis.}.
Fig.~\ref{Fig02:h1Tperp-bag-x} show results for the quantity
\ba
    h_{1T}^{\perp q}(x) = \int\di^2\vec{k}_\perp\,
    h_{1T}^{\perp q}(x,k_\perp) \label{Eq:def-h1Tperp-of-x}\;.
\ea
Bag model results for scale-dependent quantities refer to a low
initial scale $Q_0$ which is understood to be of the order of a
typical hadronic scale, say $Q_0 ={\cal O}(M_N)$, or even lower,
see e.g.\ \cite{Stratmann:1993aw}.

The $u$-quark distribution is negative. Compared to the $d$-quark
distribution, it has opposite sign to and its absolute value is four
times larger --- according to (\ref{Eq:wafe-function-SU(6)-pol}).
The pretzelosity distribution functions have opposite signs compared
to the transversity distribution functions.
Interestingly, the $h_{1T}^{\perp q}(x)$ are larger
than the transversity distribution functions $h_1^q(x)$ in the bag model,
and also larger than $f_1^q(x)$. However, it has to be noted that this
quantity, as defined in Eq.~(\ref{Eq:def-h1Tperp-of-x}), is not
constrained by positivity bounds.
The $h_{1T}^{\perp q}(x)$ reach minima or maxima, depending on the
flavor, around $x=(0.2$--$0.3)$.

What is constrained by positivity are the transverse moments
$h_{1T}^{\perp (1)q}(x)$ of the pretzelosity distribution, see
Eqs.~(\ref{Eq:positivity},~\ref{Eq:positivity-integrated},~\ref{Eq:positivity-with-h1}).
Our results for these quantities, which are defined in Eq.~(\ref{Eq:trans-mom-1}),
are shown in Fig.~\ref{Fig03:h1Tperp1-bag-x}.
The transverse moments of the pretzelosity distributions are smaller
than the transversity distributions. For $x>0.1$ their absolute values
are always more than a factor 3 smaller than those of $h_1^q(x)$.
The extrema of the  transverse moments are shifted towards larger
$x$ compared to $h_{1T}^{\perp q}(x)$, and appear
around $x=(0.3$--$0.4)$.

%

\newpage
\section{\boldmath Pretzelosity in the spectator model of \cite{Jakob:1997wg}}
\label{Sec-5:in-spectator-model}

The pretzelosity distribution was calculated in the spectator model
\cite{Jakob:1997wg}. Strictly speaking valence quark distributions were
computed there. However, in that model calculation the difference between
'quarks' and 'valence quarks' can be disregarded to a good approximation.
In  the spectator model of Ref.~\cite{Jakob:1997wg} pretzelosity is given by
\ba\label{Eq:spectator-model}
    h_{1T}^{\perp q}(x,k_\perp) = \sum_{r=s,\,a}
    L^{\! r/q} \;h_{1T}^{\perp r}(x,k_\perp)\;,\;\;\;
    L^{\! s/u} = \frac32\,,\;\;\;
    L^{\! a/u} = \frac12\,,\;\;\;
    L^{\! s/d} = 0\,,\;\;\;
    L^{\! a/d} = 1\,
\ea
where the sum goes over contributions from a scalar spectator diquark (s),
and an axial vector spectator diquark (a).
The other TMDs have the same 'decompositions' (\ref{Eq:spectator-model}),
which is dictated by the SU(6) symmetry.
The contributions of the various diquark spectators to pretzelosity, and
for comparison, the TMDs $f_1$, $g_1$ and $h_1$ read \cite{Jakob:1997wg}
\ba
    f_1^r(x,k_\perp)
    &=& \phantom{a_r}\,J_r(x,k_\perp)\,\biggl[ (xM_N+m)^2+k_\perp^2\biggr] \\
    g_1^r(x,k_\perp)
    &=& a_r \, J_r(x,k_\perp)\,\biggl[ (xM_N+m)^2-k_\perp^2\biggr] \\
    h_1^r(x,k_\perp)
    &=& a_r \, J_r(x,k_\perp)\,\biggl[ (xM_N+m)^2\,\biggr] \\
    h_{1T}^{\perp r}(x,k_\perp)
    &=& a_r \, J_r(x,k_\perp)\,\biggl[ -2 M_N^2\;\biggr]
\ea
where the 'spin factor' $a_r$ assumes the values $a_s=1$ and $a_a = -\,\frac13$,
and
\be\label{Eq:spectator-details}
    J_r(x,k_\perp) =
    \frac{N^2(1-x)^{2\alpha-1}}{2(2\pi)^3(k_\perp^2+\lambda_r^2)^{2\alpha}}
    \,,\;\;\;
    \lambda_r^2 = \Lambda^2(1-x)+xM_r^2 -x(1-x)M_N^2\;.
\ee
The meaning and the values of the parameters are:
spectator masses $M_a    =0.8 \,{\rm GeV}$
and      $M_s    =0.6 \,{\rm GeV}$,
quark mass   $m      =0.36\,{\rm GeV}$,
cutoff       $\Lambda=0.5 \,{\rm GeV}$,
and          $\alpha =2$.
The normalization factor $N$ is such that $\int\di x\,f_1^r(x)=1$, and for the
nucleon mass the rounded value $M_N=0.94\,{\rm GeV}$ was used \cite{Jakob:1997wg}.
Other choices of parameters are also possible, see \cite{Jakob:1997wg} for details.
The results refer to a scale as low as the initial scale of the GRV
parameterization $\mu_0^2=0.23\,{\rm GeV}^2$ \cite{Gluck:1998xa},
and possibly even lower \cite{Jakob:1997wg}.

The contributions from the respective diquarks satisfy the relations, for example,
\ba
    a_r\,f_1^r(x,k_\perp) + g_1^r(x,k_\perp) = 2h_1^r(x,k_\perp)
    \label{Eq:spec-relations-TMD-r-I}\\
    h_1^r(x,k_\perp) - h_{1T}^{\perp(1)r}(x,k_\perp) = a_r\,f_1^r(x,k_\perp)
    \label{Eq:spec-relations-TMD-r-II}
\ea
which are the analogs of the bag model relations
(\ref{Eq:specific-rel-I},~\ref{Eq:specific-rel-II}) discussed in
Sec.~\ref{Sec-4:pretzelosity-in-bag}. But, since $M_a\neq M_s$, the relations
(\ref{Eq:spec-relations-TMD-r-I},~\ref{Eq:spec-relations-TMD-r-II})
in general do not give rise to relations among TMDs for specific flavors.
More precisely, one recovers the relation (\ref{Eq:specific-rel-II-flavour})
in the spectator model for $d$-flavor, since here the scalar quark decouples.
For the $u$-flavor the relation (\ref{Eq:specific-rel-II-flavour})
follows only in the limit $M_a\to M_s$, because in that limit
the decompositions (\ref{Eq:spectator-model}) coincide with
the bag model SU(6) flavor relations
(\ref{Eq:wafe-function-SU(6)-pol},~\ref{Eq:wafe-function-SU(6)-unp}).
Therefore, although both are based on the SU(6) symmetry,
the spectator model is more general than the bag model
with respect to the flavor dependence.

Notice that in spectator models the mass difference between scalar
and axial-vector diquarks is responsible for the $\Delta$-nucleon
mass splitting whose physical value is reproduced for
$M_a-M_s=200\,{\rm MeV}$ \cite{Close:1988br}. In the large-$N_c$ limit
we have $M_\Delta-M_N={\cal O}(N_c^{-1})$, which implies that also
$M_a-M_s\to 0$ in that limit. As for $M_a=M_s$ the flavor relations
(\ref{Eq:wafe-function-SU(6)-pol},~\ref{Eq:wafe-function-SU(6)-unp}),
are recovered, see above, the same discussion applies here as in
Sec.~\ref{Sec-4:pretzelosity-in-bag}. Thus, the spectator model
respects the large-$N_c$ relation (\ref{Eq:large-Nc}) ---
formally and practically, although with non-negligible
$1/N_c$-corrections, see below.

From this discussion we learn an interesting lesson about the model
dependence of relations among different (in QCD independent!) TMDs.
Relations that involve unpolarized and polarized TMDs are
'flavor dependent', see
(\ref{Eq:specific-rel-I-flavour},~\ref{Eq:specific-rel-II-flavour}),
and valid only in 'flavor-blind' models like the bag model,
see Sec.~\ref{Sec-4:pretzelosity-in-bag} and \cite{Barone:2001sp},
or constituent quark models, see \cite{Pasquini:2005dk}.
Such relations break down in general in models with non-trivial flavor
structure. In fact, the different masses of scalar and axial-vector diquarks
in the spectator model spoil such relations.

What about relations involving polarized TMDs only? If we subtract the relations
(\ref{Eq:spec-relations-TMD-r-I},~\ref{Eq:spec-relations-TMD-r-II}) from each
other $f_1^r(x,k_\perp)$ drops out, and since in that model the flavor structure
is precisely the same for all polarized TMDs (independently of the scalar and
axial-vector diquark masses), we recover Eq.~(\ref{Eq:measure-of-relativity}).
Thus, the relation (\ref{Eq:measure-of-relativity}) among helicity,
transversity and pretzelosity seems to be valid in a larger class of models.
We shall come back to this point in Sec.~\ref{Sec-6:validity-of-relation}.

\begin{figure}[b]
        \includegraphics[width=7.5cm]{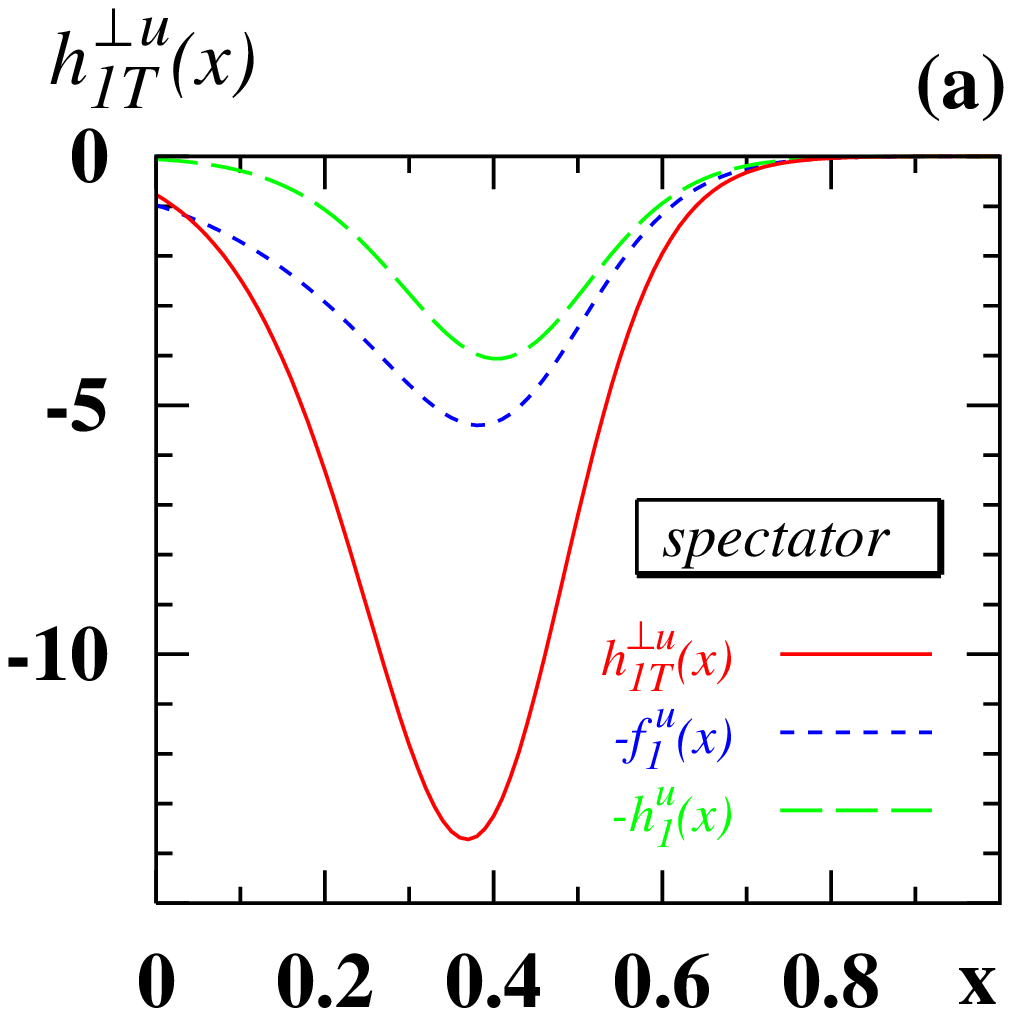}
        \includegraphics[width=7.5cm]{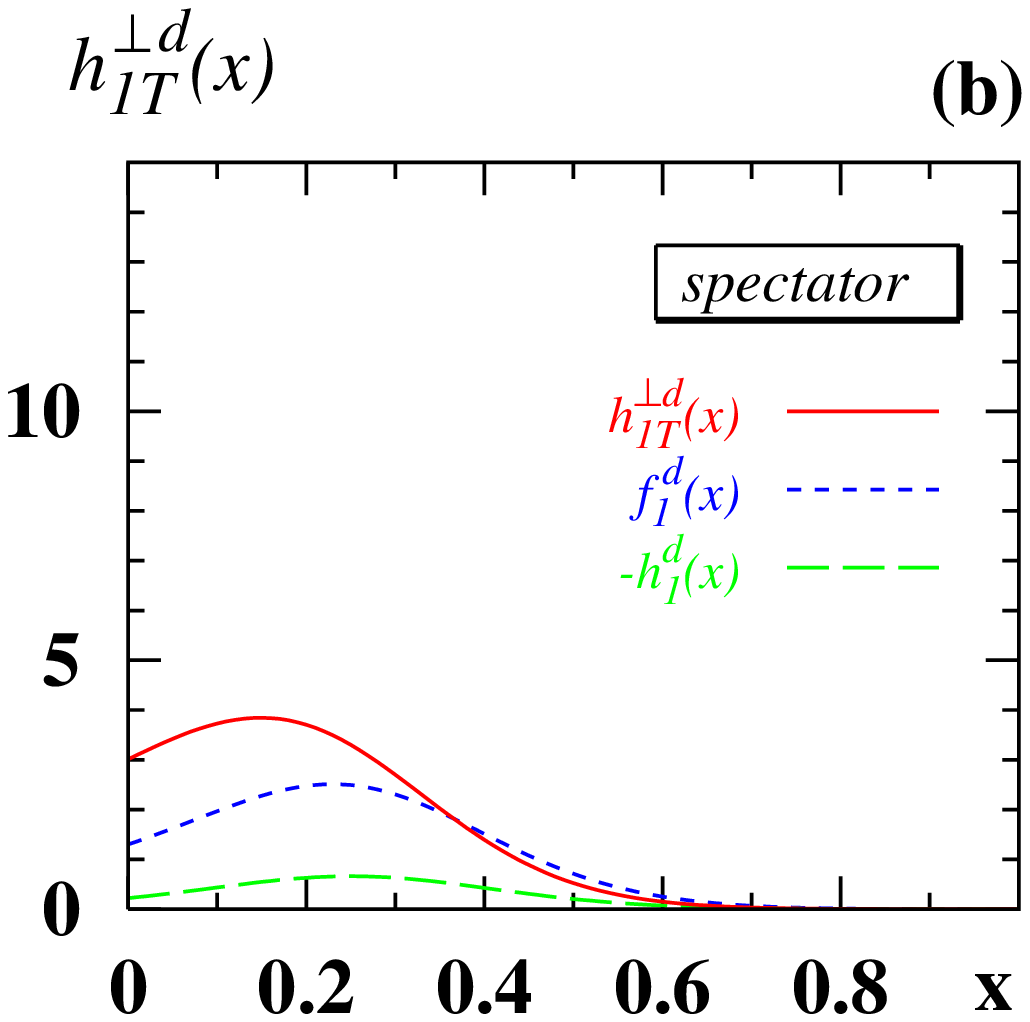}
\caption{\label{Fig04:spec-h1Tperp-x}
    The parton distribution function $h_{1T}^{\perp q}(x)$ vs.\ $x$
    from the spectator model of Ref.~\cite{Jakob:1997wg} in comparison
    to $f_1^q(x)$ and $h_1^q(x)$ from the same model.
    The functions $h_{1T}^{\perp q}(x)$ are rather large, even larger
    than $f_1^q(x)$. Notice, however, that $h_{1T}^{\perp q}(x)$ itself,
    as defined in (\ref{Eq:def-h1Tperp-of-x}), is not constrained
    by positivity bounds. All results refer to the low scale of the
    model \cite{Jakob:1997wg}.}

        \includegraphics[width=7.5cm]{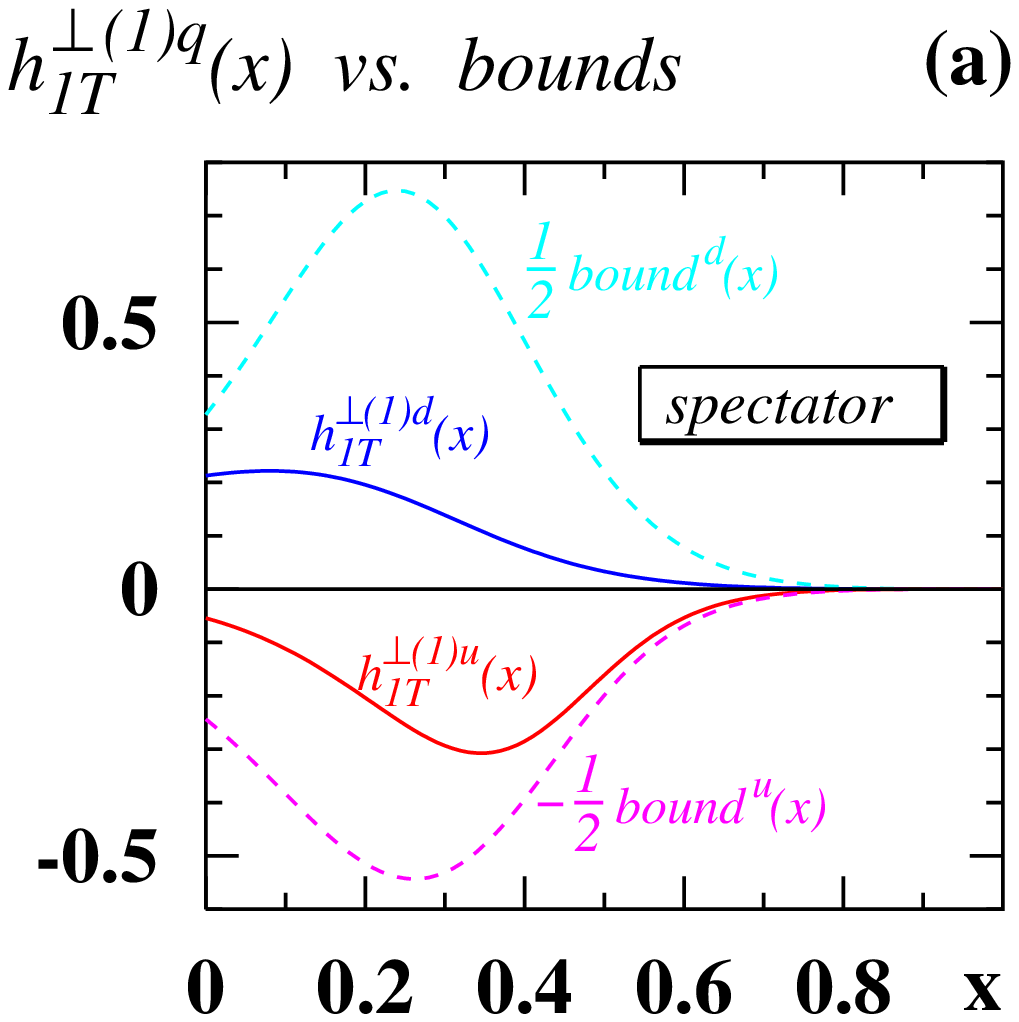}
    \includegraphics[width=7.5cm]{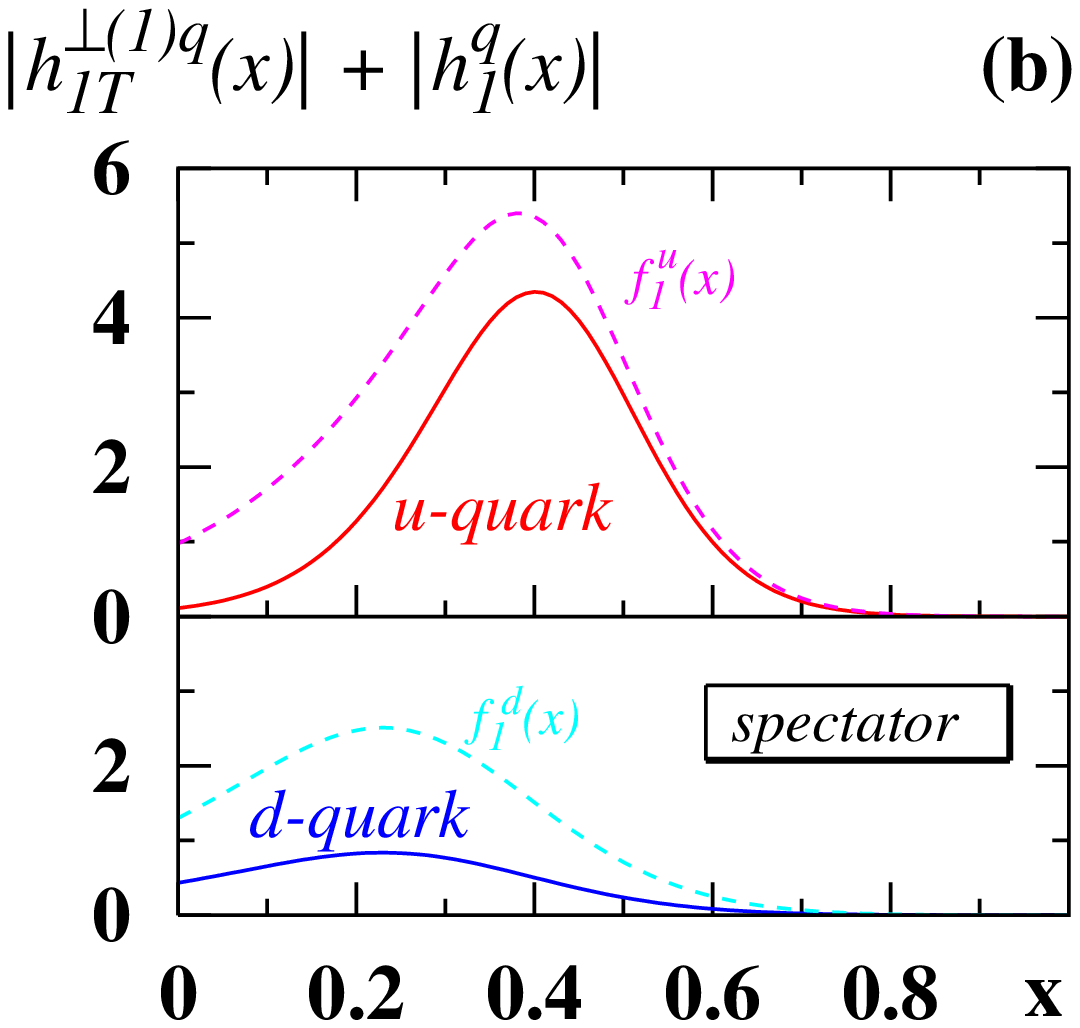}
\caption{\label{Fig05:spec-h1Tperp1-x}
    (a) The transverse moment $h_{1T}^{\perp (1)q}(x)$ vs.\ $x$
    (solid lines). It does not exceed even half of the
    bound$^q(x)\equiv \frac12(f_1^q-g_1^q)(x)$ (dashed lines)
    allowed by positivity, see Eq.~(\ref{Eq:positivity-integrated}).
    (b) The combination $|h_{1T}^{\perp (1)q}(x)|+|h_1^q(x)|$
    as function of $x$ (solid lines). This combination
    must not exceed $f_1^q(x)$ (dashed lines),
    see Eq.~(\ref{Eq:positivity-with-h1}), which is the case.}
\end{figure}

Let us now discuss numerical results. Fig.~\ref{Fig04:spec-h1Tperp-x} shows
$h_{1T}^{\perp q}(x)$ defined in (\ref{Eq:def-h1Tperp-of-x}). The $u$-quark
distribution is negative and (in magnitude) much larger than the positive
$d$-quark distribution. The $h_{1T}^{\perp q}(x)$ are larger than $h_1^q(x)$
in that model, and also larger than $f_1^q(x)$ (we recall that this quantity
is not constrained by positivity).

Fig.~\ref{Fig05:spec-h1Tperp1-x}a shows the transverse moments
$h_{1T}^{\perp (1)q}(x)$.
One can convince oneself that the positivity constraints
(\ref{Eq:positivity-integrated},~\ref{Eq:positivity-with-h1})
are always satisfied in the model of \cite{Jakob:1997wg}
independently of the specific values for the spectator masses.
However, here the flavor structure is more involved, such that it is not
possible to see a priori what fraction of the positivity bound the transverse
moment of pretzelosity explores --- in contrast to bag model,
Eq.~(\ref{Eq:specific-ineq}) --- and it is worth to look at this
in more detail.
Fig.~\ref{Fig05:spec-h1Tperp1-x}a shows that the transverse moments of
pretzelosity never exceed even half of the (\ref{Eq:positivity-integrated}).
Also the positivity condition (\ref{Eq:positivity-with-h1}) is satisfied,
see Fig.~\ref{Fig05:spec-h1Tperp1-x}b. Noteworthy, pretzelosity and
transversity use, especially for the $u$-flavor, a large fraction
of the room allowed by positivity conditions.

\begin{figure}[b]
        \includegraphics[width=8cm]{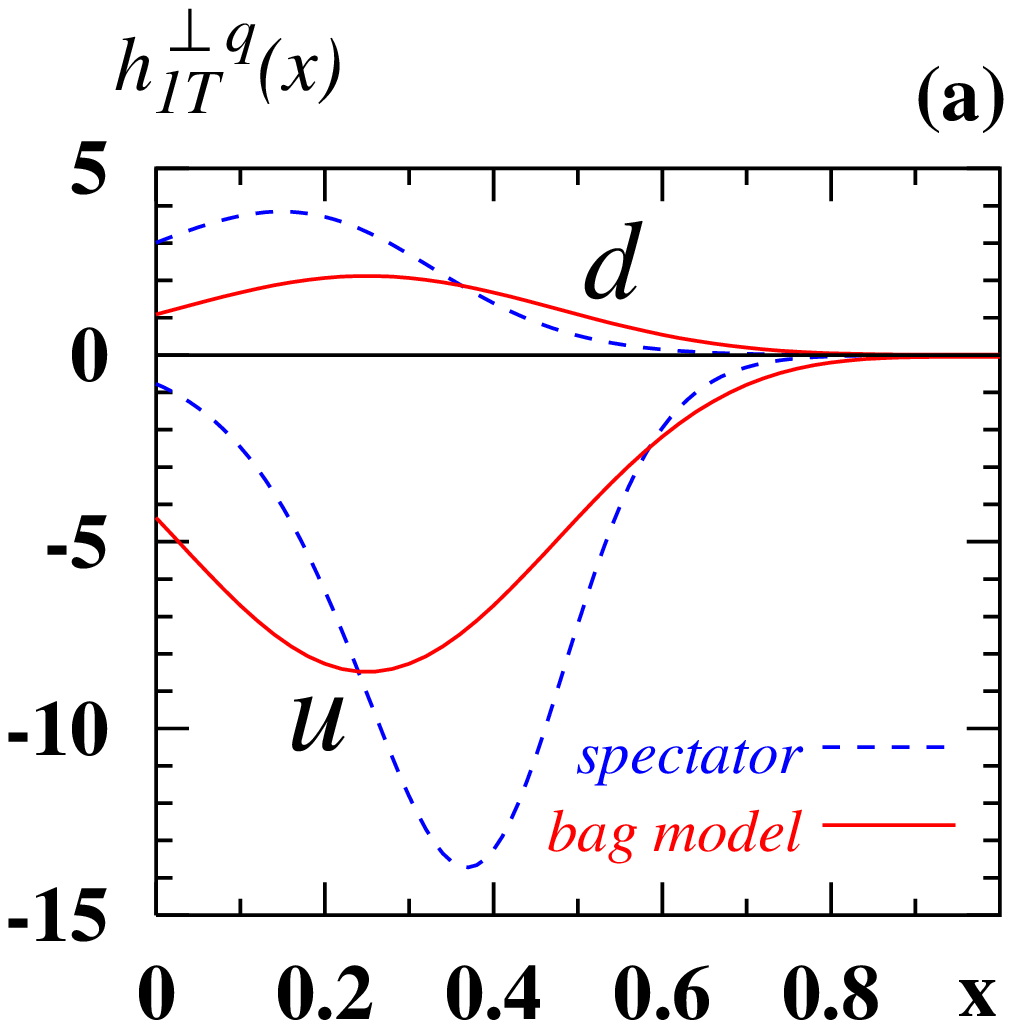}
    \includegraphics[width=8cm]{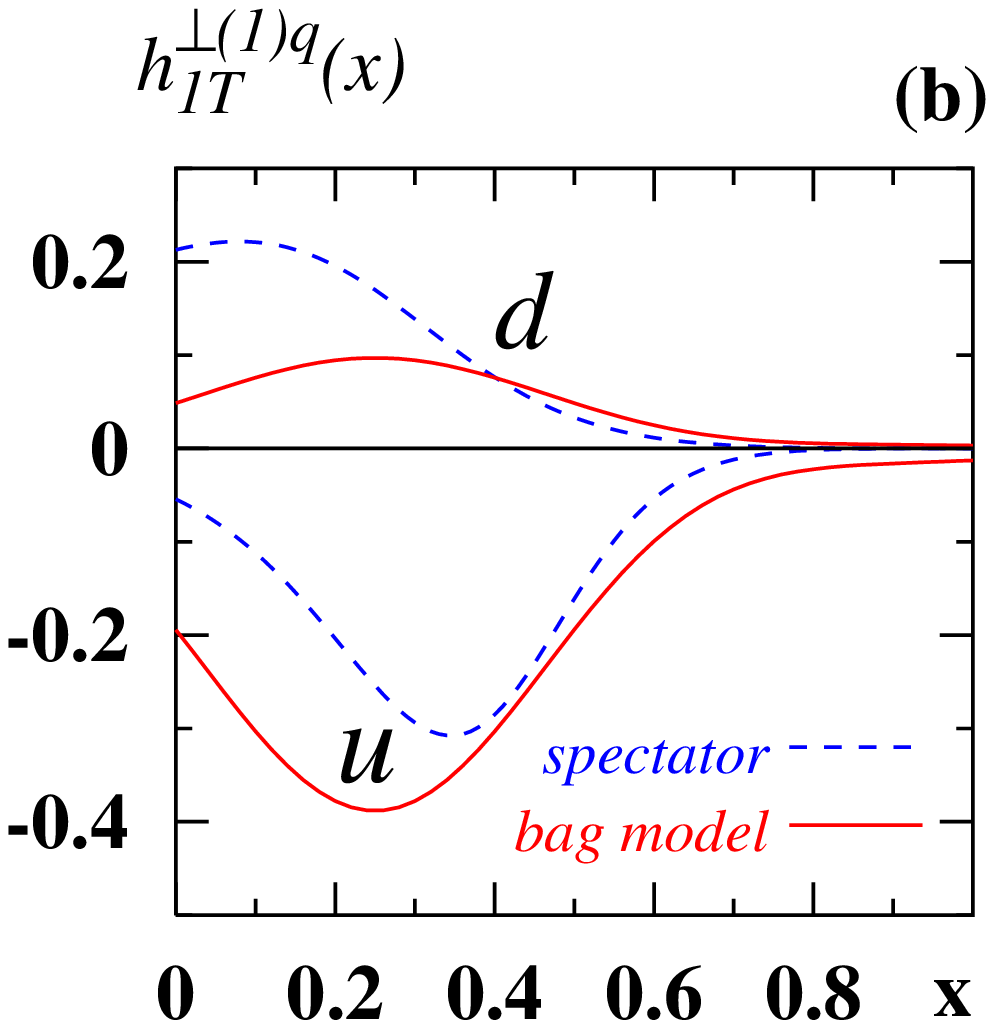}

        \includegraphics[width=8cm]{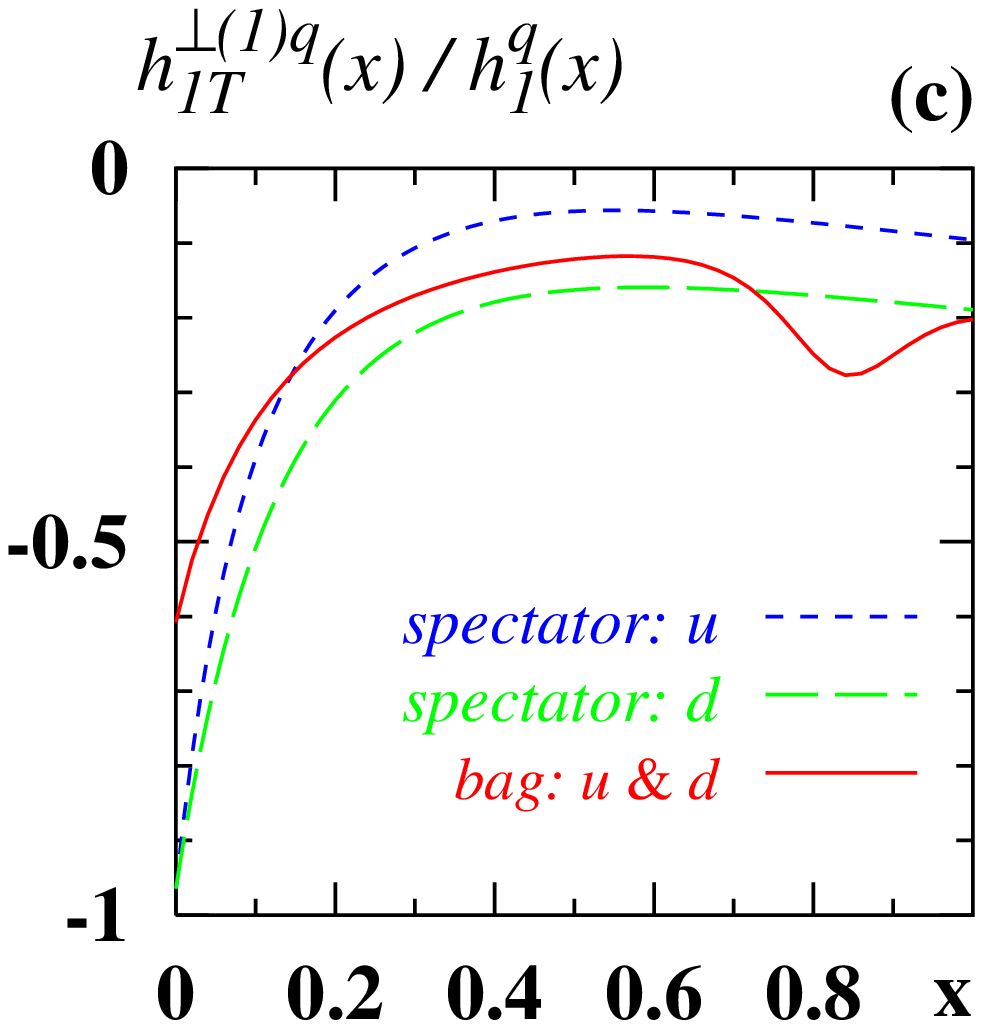}
    \includegraphics[width=8cm]{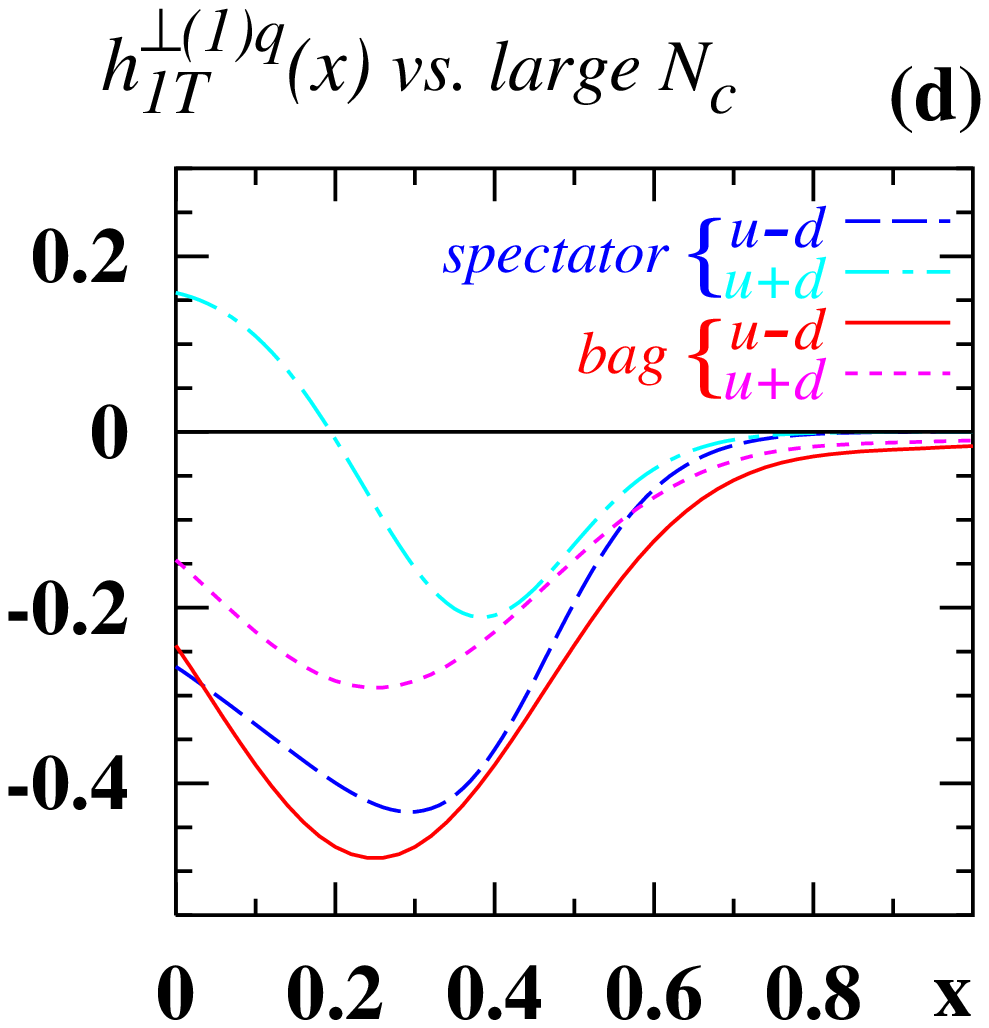}
\caption{\label{Fig06:bag-vs-spec}
    Comparison of results from bag model (computed here, cf.\
    Sec.~\ref{Sec-4:pretzelosity-in-bag}) and spectator model,
    Ref.~\cite{Jakob:1997wg}.
    (a) $h_{1T}^{\perp q}(x)$ vs.~$x$.
    (b) $h_{1T}^{\perp (1)q}(x)$ vs.~$x$.
    (c) The ratio of $h_{1T}^{\perp (1)q}(x)$ to $h_1^a(x)$ vs.~$x$.
    (d) The flavor combinations $(h_{1T}^{\perp (1)u}\pm h_{1T}^{\perp (1)d})(x)$
    vs.\ $x$.
    All results refer to the low scales of these models.}
\end{figure}

Next we confront the results from the bag model (obtained here,
see Sec.~\ref{Sec-4:pretzelosity-in-bag}) and the spectator model
of Ref.~\cite{Jakob:1997wg}.
There is a good qualitative agreement, the signs and magnitudes
of $h_{1T}^{\perp q}(x)$ and  $h_{1T}^{\perp (1)q}(x)$ agree,
see Fig.~\ref{Fig06:bag-vs-spec}a and b.
The transverse moments $h_{1T}^{\perp (1)q}(x)$ are much smaller
than the transversity distributions $h_1^q(x)$.
With the exception of very small-$x$ they are at least 3 times smaller
than $h_1^q(x)$, see Fig.~\ref{Fig06:bag-vs-spec}c.

Let us finally compare the flavor combinations
$(h_{1T}^{\perp (1)u}\pm h_{1T}^{\perp (1)d})(x)$ in
Fig.~\ref{Fig06:bag-vs-spec}d. As we have discussed,
above and in Sec.~\ref{Sec-4:pretzelosity-in-bag},
both models are conceptually in agreement with large-$N_c$.
But in both models, of course, the finite value $N_c=3$ is used.
It is therefore of interest to investigate to which extent the
models reflect the large-$N_c$ pattern in Eq.~(\ref{Eq:large-Nc}).
In fact, Fig.~\ref{Fig06:bag-vs-spec}d shows that the modulus of
the $(u-d)$ flavor combination is always larger than that of the
$(u+d)$ flavor combination, as one would expect on the basis of
large-$N_c$ arguments in Eq.~(\ref{Eq:large-Nc}).
But the difference between the magnitudes of the $(u\pm d)$ flavor combinations
is not pronounced. This is true especially in the bag model where
the 'large' $(u-d)$ flavor combination is only $\frac{5}{3}$-times larger than
the 'small' $(u+d)$ flavor combination. (Recall that this factor is
$\frac{N_c+2}{3}$, i.e.\ in fact large for $N_c\to\infty$.)
The spectator model reflects more clearly the large-$N_c$ behavior predicted
in Eq.~(\ref{Eq:large-Nc}) --- especially if we recall that these predictions
are valid for $x\sim {\cal O}(1/N_c)$ \cite{Pobylitsa:2003ty}.

To which extent can one expect models formulated for finite $N_c=3$
to respect large-$N_c$ predictions of the kind (\ref{Eq:large-Nc})?
Fig.~\ref{Fig06:bag-vs-spec}d gives a flavor on that. In nature,
large-$N_c$ results are found to hold within a similar accuracy
\cite{Efremov:2000ar}. Interestingly, the Sivers function predicted
to exhibit in the large-$N_c$ limit a flavor dependence analog to
(\ref{Eq:large-Nc}), see \cite{Pobylitsa:2003ty}, seems to obey
large-$N_c$ predictions rather closely
\cite{Efremov:2004tp,Collins:2005ie,Collins:2005rq,Vogelsang:2005cs,Anselmino:2005nn,Arnold:2008ap,Anselmino:2008sg}.

\newpage
\section{\boldmath How general is the relation in Eq.~(\ref{Eq:measure-of-relativity})}
\label{Sec-6:validity-of-relation}

In the bag and spectator model the relation
(\ref{Eq:measure-of-relativity}) is valid not only for quark but
also for antiquark distributions. Hereby one has to take into
account that the Dirac-structure in the chirally odd $h_1$ and
$h_{1T}^\perp$ is C-odd, while it is C-even for $g_1$. Thus, for
antiquarks the relation (\ref{Eq:measure-of-relativity}) reads
$g_1^a(x,k_\perp) + h_1^a(x,k_\perp) = -
h_{1T}^{\perp(1)a}(x,k_\perp)$ with $a=\bar u$, $\bar d$.

While strictly speaking in the bag model anti-quark distributions are unphysical,
see footnote~\ref{footnote:qbar-in-bag-model}, one obtains in principle consistent
results in the spectator model. In \cite{Jakob:1997wg} no antiquark distributions
were considered, however, this can be done precisely as sketched in
Eqs.~(\ref{Eq:spectator-model}-\ref{Eq:spectator-details}) but with additional
parameters for the (four-quark-) spectators, which must be tuned.
The results, however, are analog to
(\ref{Eq:spectator-model}-\ref{Eq:spectator-details}) and satisfy the
relation (\ref{Eq:measure-of-relativity}) for antiquarks.

So, in the quark models we have discussed the relation
(\ref{Eq:measure-of-relativity}) holds for both, quark and
antiquark TMDs. It will also be valid in a large class of
relativistic models, e.g.\ the constituent quark models of the
kind used in \cite{Pasquini:2005dk}, or in the model of
\cite{Efremov:2004tz} (which remains to be verified by direct
calculations). The question that naturally arises is: How general
is the relation (\ref{Eq:measure-of-relativity})?

When discussing this question it is important to observe that
Eq.~(\ref{Eq:measure-of-relativity}) relates chirally even and chirally odd TMDs,
i.e.\ it relates quark (and anti-quark) TMDs having gluon 'partners' and such
having no gluon counterparts, cf.\ Sec.~\ref{Sec-3:What-we-know}.
It is clear that the evolution properties of these TMDs are different.
Therefore, if the relation (\ref{Eq:measure-of-relativity})  were valid at
some initial scale, then it certainly would not be valid at a different scale.

These considerations imply, that the relation (\ref{Eq:measure-of-relativity})
can be valid only in 'no-gluon models'.
This expectation is supported by the model calculations of Ref.~\cite{Meissner:2007rx}.
There (among others) quark and gluon TMDs in a hypothetical 'quark target' were computed.
It was found that $g_1(x,k_\perp)-h_1(x,k_\perp)\neq 0$ (meaning that the model
is 'relativistic') but difference was not related to pretzelosity.
The latter is zero in that model
(to the considered order of '$\alpha_s$') \cite{Meissner:2007rx}.
This is in line with our expectations: the explicit inclusion
of gluon degrees of freedom spoils (\ref{Eq:measure-of-relativity}).

Nevertheless, Eq.~(\ref{Eq:measure-of-relativity}) could turn out
to be approximately satisfied, and useful for estimating
pretzelosity on the basis of the helicity and transversity
distributions, which we know presently better, or are on a good
way to that
\cite{Vogelsang:2005cs,Efremov:2006qm,Anselmino:2007fs}. At this
point, since this relation is found in models without gluons
(where we simply could 'neglect' the gauge-links), presumably the
$k_\perp$-integrated version of (\ref{Eq:measure-of-relativity})
is a more reliable prediction of relativistic quark models.

\section{\boldmath Preliminary COMPASS data, and future experiments at JLab}

\begin{figure}[t!]
\begin{tabular}{cc}
    \hspace{-0.5cm}\includegraphics[width=7.5cm]{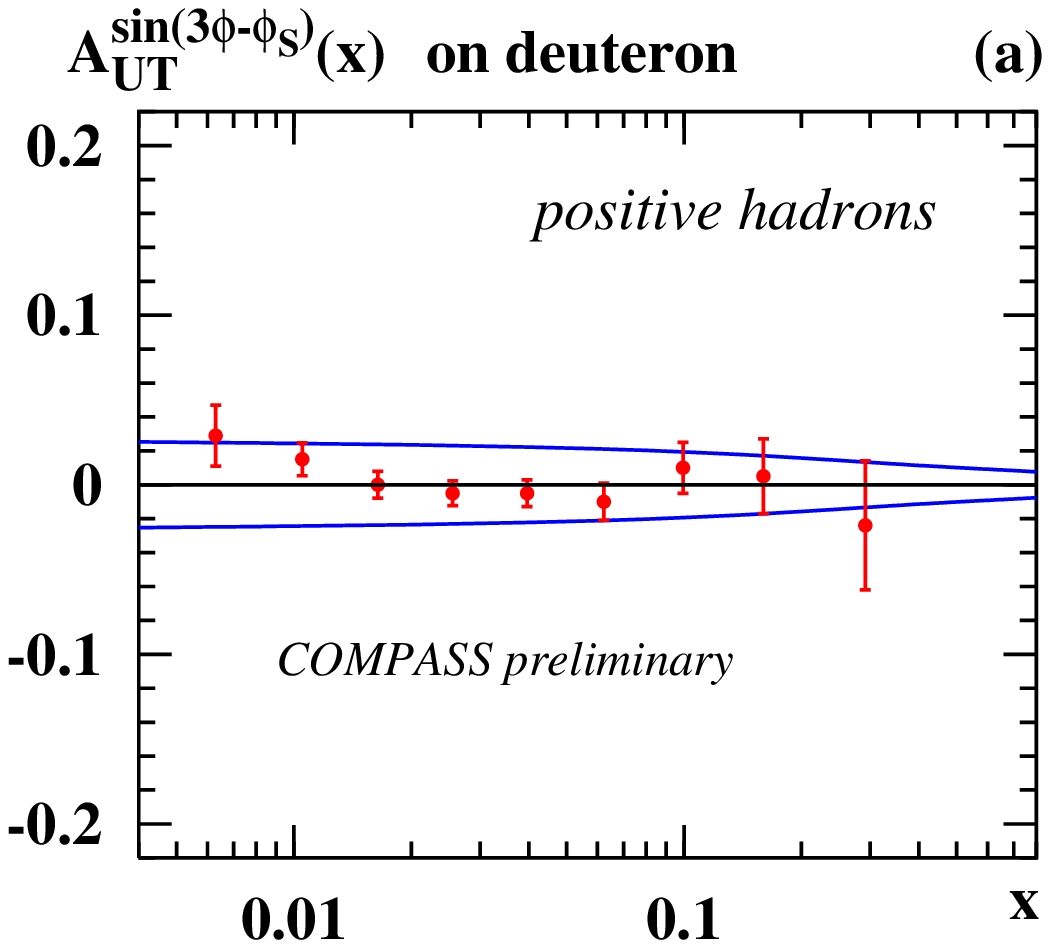} &
    \hspace{-0.5cm}\includegraphics[width=7.5cm]{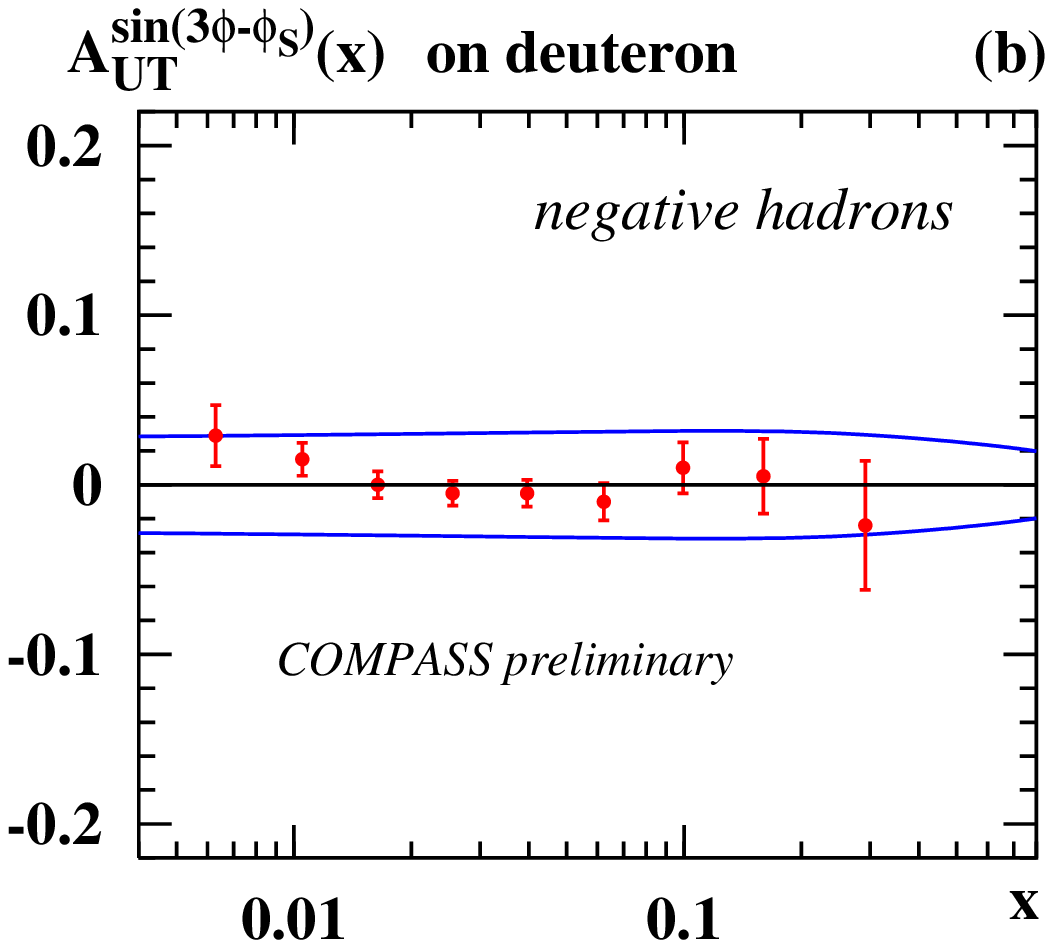}
\end{tabular}
\caption{\label{Fig03:AUT-deut-COMPASS-x}
    The transverse target SSA $A_{UT,\,\pi}^{\sin(3\phi-\phi_S)}$
    for deuteron estimated on the basis of the positivity bound vs.\
    preliminary COMPASS data \cite{Kotzinian:2007uv}.}
\end{figure}

In the COMPASS experiment the $\sin(3\phi-\phi_S)$ and other SSAs
were measured on a deuteron target \cite{Kotzinian:2007uv}.
We shall estimate the deuteron distribution functions
by neglecting nuclear binding effects and exploring
isospin symmetry as (analog for $\bar u$ and $\bar d$,
and we neglect strange and heavier quarks):
\ba
    \left.
    {h_{1T}^{\perp u/D} = h_{1T}^{\perp u/p} + h_{1T}^{\perp u/n}}\atop
    {h_{1T}^{\perp d/D} = h_{1T}^{\perp d/p} + h_{1T}^{\perp d/n}}
    \right\}            = h_{1T}^{\perp u} + h_{1T}^{\perp d}\,.
\ea
Let us denote by $N(\pi^+,D)$ the numerator of the SSA
$A_{UT}^{\sin(3\phi-\phi_S)}$ in production of positive
pions from the deuteron target. It is given by
(we skip prefactors irrelevant for the qualitative discussion)
\ba
&&  N(\pi^+,D) =
     (h_{1T}^{\perp u} + h_{1T}^{\perp d})
     (4 H_1^{\perp\rm fav}+H_1^{\perp\rm unf})
    +(h_{1T}^{\perp\bar u} + h_{1T}^{\perp\bar d})
     ( H_1^{\perp\rm fav}+4H_1^{\perp\rm unf})
\ea
We estimate the maximum effect for this SSA as follows
(and analogously for other pions)
\ba
&&  |N(\pi^+,D)|
    \le
     |h_{1T}^{\perp u} + h_{1T}^{\perp d}|\cdot
     |4 H_1^{\perp\rm fav}+H_1^{\perp\rm unf}|
    +|h_{1T}^{\perp\bar u} + h_{1T}^{\perp\bar d}|\cdot
     |H_1^{\perp\rm fav}+4H_1^{\perp\rm unf}|
\ea
where (and analogously for antiquarks)
\be\label{Eq:h1Tperp-bound-deut}
    |h_{1T}^{\perp(1) u} + h_{1T}^{\perp(1) d}| \le
    |h_{1T}^{\perp(1) u}|+|h_{1T}^{\perp(1) d}| \le
    \frac12(f_1^u+f_1^d-g_1^u-g_1^d)   \,.
\ee
Using information on Collins effect \cite{Efremov:2006qm,Anselmino:2007fs}, the
parameterizations \cite{Gluck:1998xa,Kretzer:2000yf}
for $f_1^a(x)$, $g_1^a(x)$, $D_1^a(z)$ at a $Q^2=2.5\,{\rm GeV}^2$,
and assuming that positive (negative) hadrons at COMPASS are
mainly positive (negative) pions (this is legitimate to a good approximation),
one obtains the results shown in Fig.~\ref{Fig03:AUT-deut-COMPASS-x}.
We compare the positivity-bound result to the preliminary data \cite{Kotzinian:2007uv}.

%
%

At larger $x>0.1$ the statistics analyzed in \cite{Kotzinian:2007uv}
is not sufficient to be sensitive to the pretzelosity distribution.
However, below $x<0.1$ COMPASS data favor that pretzelosity does
not saturate its bound. (This would be in line with our models. But those
refer to low scales, and are strictly speaking not applicable to small-$x$.)

This could hint at the following. Either, the chiral-odd pretzelosity is
suppressed at small-$x$ with respect to the chiral-even $f_1$, which is
expected, see Sec.~\ref{Sec-3:What-we-know}.
Or, the $u$ and $d$-flavors have opposite signs, as predicted in the
large-$N_c$ limit and the models. Or, both. In view of the early stage
of art 
it is too early to draw definite conclusions.

An important observation, however, is that in spite
of the small (consistent with zero) effect observed at COMPASS,
pretzelosity does not need to be small in the region $x>0.1$,
see Fig.~\ref{Fig03:AUT-deut-COMPASS-x},
where JLab can measure with great precision.
So there is no discouragement due to small COMPASS pretzelosity effect
for the planned CLAS measurements!

It remains to be mentioned that HERMES also studied the
$\sin(3\phi-\phi_S)$ azimuthal modulation in the transverse target experiments.
It was quoted that this SSA is zero within error bars, see e.g.\
\cite{Airapetian:2004tw,Diefenthaler:2005gx,Diefenthaler:2007rj}.
Presumably, an SSA as large as what we obtain from saturating positivity
(see below) would have been seen at HERMES. However, since no HERMES
data were shown on that observable, we cannot draw quantitative
conclusions from that.

%
\begin{figure}[b!]
\begin{tabular}{ccc}
    \hspace{-0.5cm}\includegraphics[width=6cm]{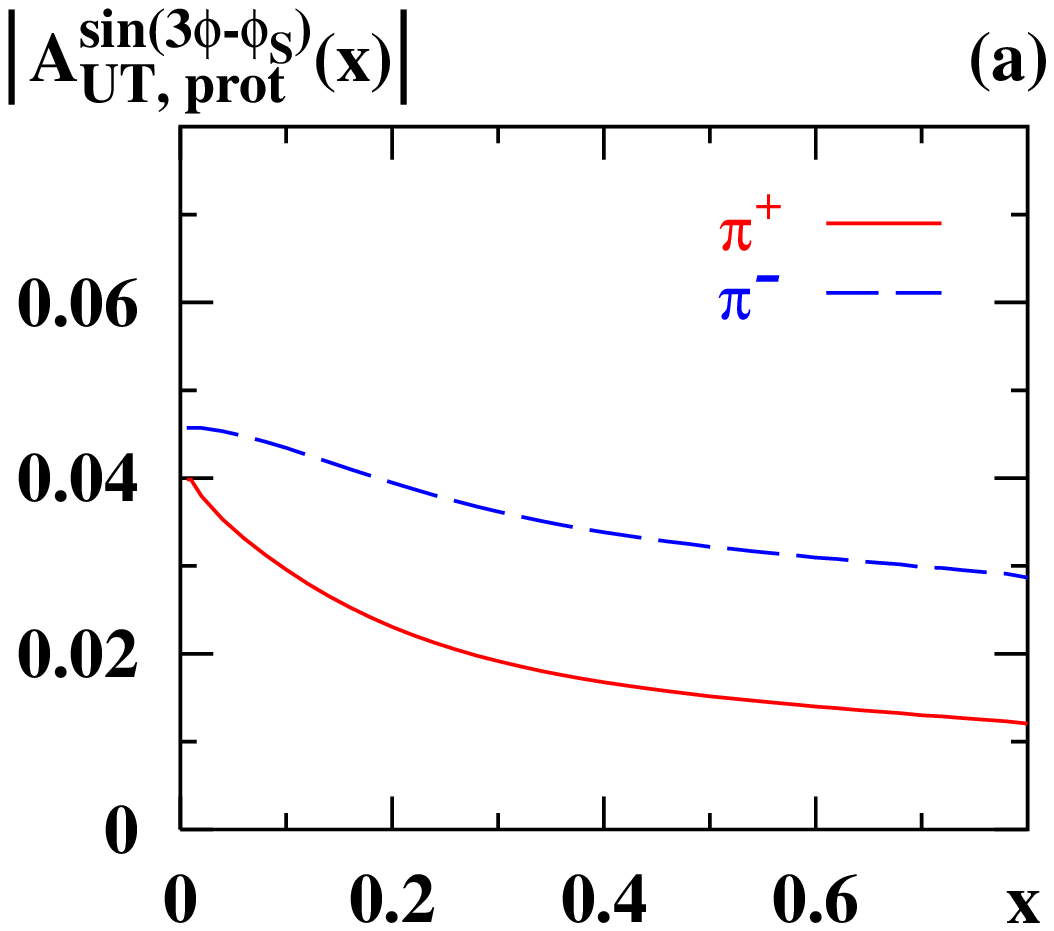} &
    \hspace{-0.5cm}\includegraphics[width=6cm]{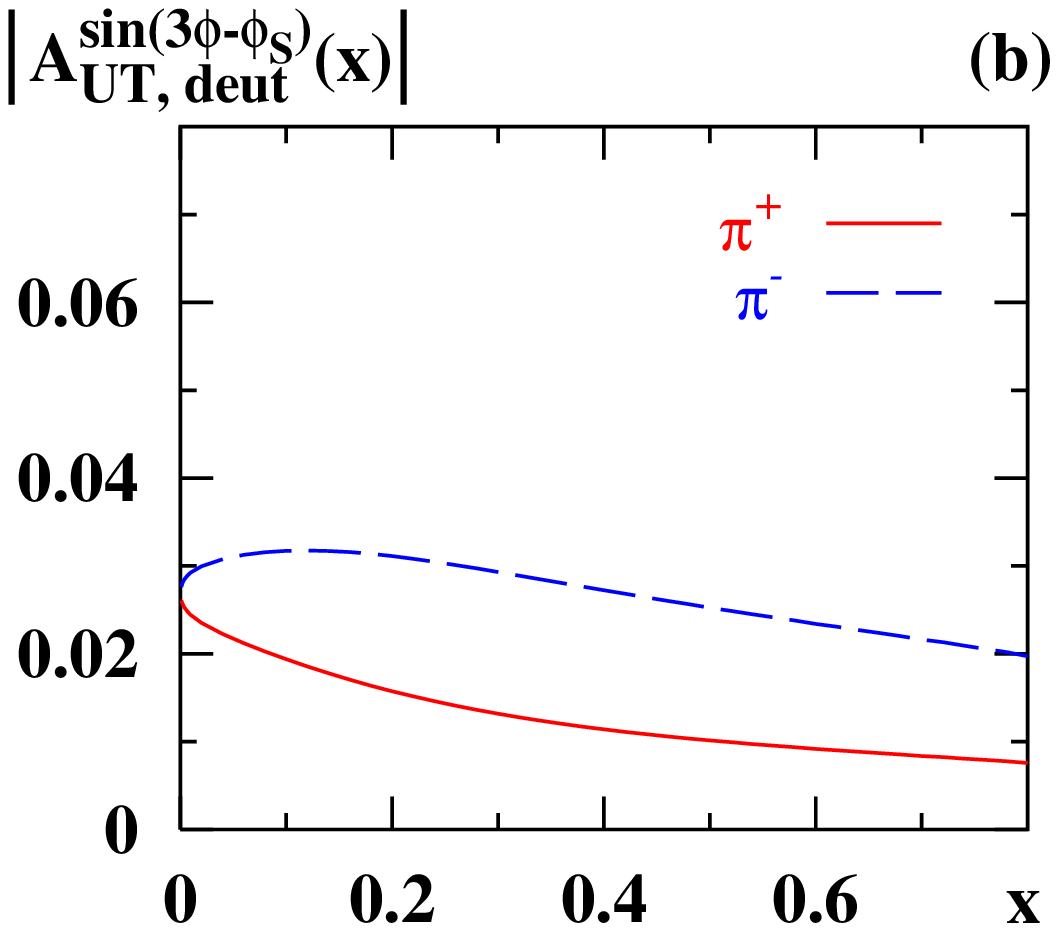} &
    \hspace{-0.5cm}\includegraphics[width=6cm]{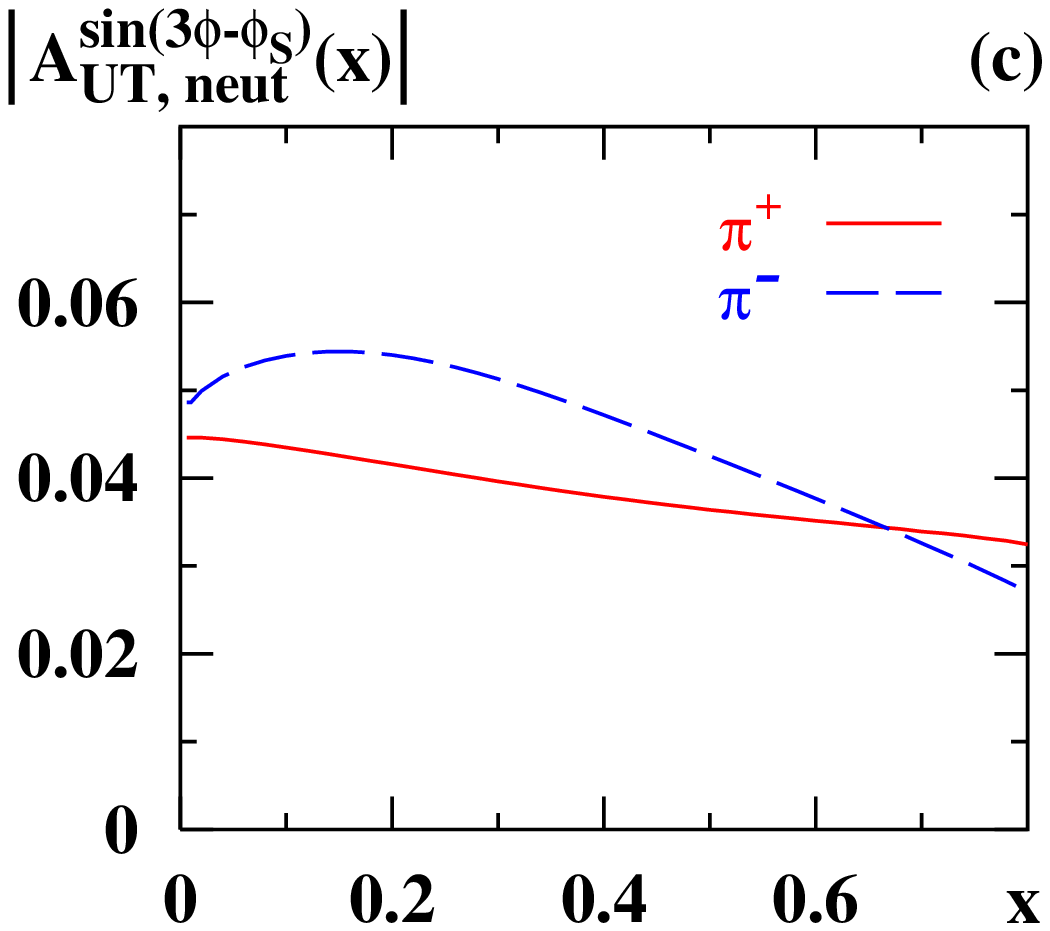}
\end{tabular}
\caption{\label{FigXX:AUT-various-targets}
    The modulus of the transverse target SSA
    $A_{UT}^{\sin(3\phi-\phi_S)}$ in charged pion production from proton,
    deuteron, and neutron targets as function of $x$. Estimates on the
    basis of the positivity bound.}
\end{figure}

In order to see how further analyses from HERMES, as well as future experiments
at COMPASS and Jefferson Lab could improve our understanding of pretzelosity,
let us estimate the upper bounds for the modulus of the SSA for
charged pion production from various targets. Using the positivity bound
(\ref{Eq:positivity-integrated}) we obtain the results in
Fig.~\ref{FigXX:AUT-various-targets}.

\begin{figure}[t!]
\vspace{-3mm}
    \includegraphics[width=12cm]{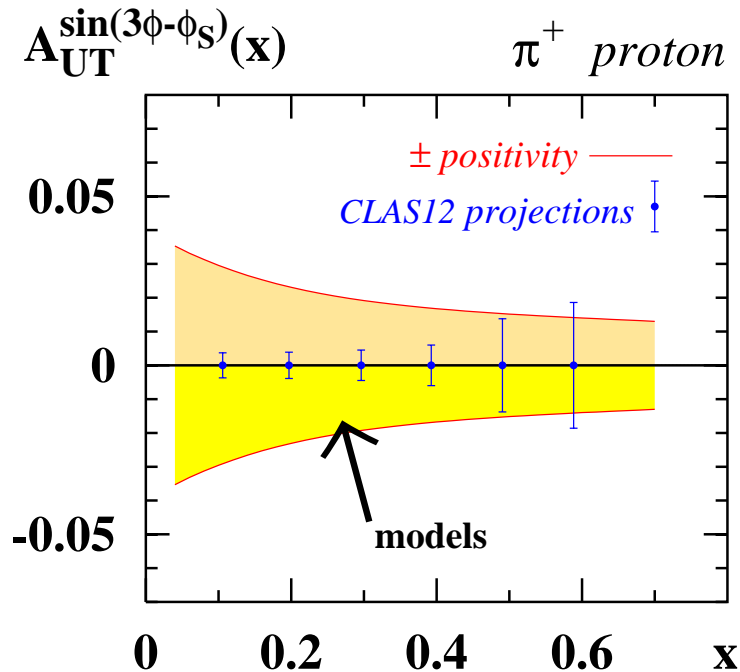}
\caption{\label{Fig9:pretzel-at-CLAS}
    The SSA $A_{UT}^{\sin(3\phi-\phi_S)}$ in $\pi^+$ production
    from a proton target in the kinematics of CLAS 12 as function of $x$.
    The error projections are from \cite{Avakian-LOI-CLAS12}.
    The shaded areas indicate the range allowed by positivity bounds.
    The bag model (obtained here) and spectator model \cite{Jakob:1997wg}
    predict a negative sign for the SSA, i.e.\  in the lower shaded area.}
\end{figure}

The SSAs for charged pions could reach up to $\sim 5\%$.
Even asymmetries reaching half or one third of that size would be measured,
especially in the region of $x \sim(0.1$--$0.4)$ in the CLAS experiment.
This is shown in Fig.~\ref{Fig9:pretzel-at-CLAS}, where we plot the $\pi^+$
SSA from a proton target in the kinematics of CLAS with 12 GeV beam upgrade.
Shown are also error projections for 2000 hours run time from
\cite{Avakian-LOI-CLAS12}. Notice that the models predict a negative SSA.

We add that the $\pi^0$ SSA is small on any target, because
the modulus of this SSA is proportional to
$|H_1^{\perp\rm fav}+H_1^{\perp\rm unf}|$
which is small (actually zero within error bars),
since $H_1^{\perp\rm unf} \approx - H_1^{\perp\rm fav}$
holds within error bars
\cite{Vogelsang:2005cs,Efremov:2006qm,Anselmino:2007fs}.

%


\section{Conclusions}

We reviewed and discussed the properties of the pretzelosity distribution function
$h_{1T}^{\perp}$, and presented a study of this leading-twist, chiral-odd,
transverse parton momentum dependent distribution function in the bag model,
and supplemented our findings with a detailed comparison to spectator model results
\cite{Jakob:1997wg}.

In these models we observed an interesting relation, which we expect to be valid
at low scales in a wide class of relativistic models. It can be summarized for
illustrative purposes by the following assertion:
\be\label{Eq:relation-illust}
    \mbox{helicity} - \mbox{transversity} = \mbox{pretzelosity}.
\ee
That the difference between the helicity and transversity distributions is
'a measure of relativistic effects' is known since long ago \cite{Jaffe:1991ra}
(and was also recognized in a bag model calculation).
However, now we are in a position to make this statement more precise.
This difference is just pretzelosity. Thus, $h_{1T}^\perp$ 'measures'
relativistic effects in the nucleon, and vanishes in the non-relativistic
limit where helicity and transversity distributions become equal.

This relation is not supported in models with explicit gluon degrees of
freedom \cite{Meissner:2007rx}, and, of course, cannot be true in QCD where
all (eight) transverse momentum dependent parton distribution functions
are linearly independent.
Nevertheless, the relation (\ref{Eq:relation-illust}), see
Eq.~(\ref{Eq:measure-of-relativity}) for its precise formulation,
could turn out to be a useful approximation.
In view of the numerous novel functions involved, any well-motivated
approximation is welcome and valuable \cite{Avakian:2007mv}.

Besides being useful for extending our intuition on relativistic spin-orbit
effects in nucleon \cite{Miller:2007ae,Burkardt:2007rv}, the relation
(\ref{Eq:relation-illust}) has also an important consequence on transversity.
In the bag and spectator model  $h_{1T}^{\perp u}$ is negative. Since
$g_1^u(x)$ is positive, this implies that $h_1^u(x) > g_1^u(x)$. For the $d$-flavor
signs are reversed, but in any case $|h_1^q(x)|>|g_1^q(x)|$. This is found
also in models, e.g.\  \cite{Schweitzer:2001sr}.

In the bag model, the negative sign of $h_{1T}^{\perp u}$ arises
because it is proportional to {\sl minus} the square of the
$p$-wave component of the nucleon wave function. Thus, in models
with no higher orbital momentum ($d$-wave, etc.) components,
$h_{1T}^{\perp u}$ is manifestly negative ($h_{1T}^{\perp d}$ has
opposite sign dictated by SU(6) symmetry, and predicted in large
$N_c$ \cite{Pobylitsa:2003ty}).

This prediction can be tested at JLab. Since the production
of positive pions from a proton target is dominated by the $u$-flavor,
one expects a negative $\sin(3\phi-\phi_S)$ SSA, see Fig.~\ref{Fig9:pretzel-at-CLAS}.

Forthcoming analyses and experiments at COMPASS, HERMES and JLab
\cite{Avakian-LOI-CLAS12,Avakian-clas6,Avakian-clas12}
will provide valuable information on the pretzelosity distribution
function, and deepen our understanding of the nucleon structure.

\vspace{0.5cm}

\noindent{\bf Acknowledgements.}
We thank Stephan Meissner for discussions.
The work is partially supported by BMBF (Verbundforschung),
German-Russian collaboration (DFG-RFFI) under contract number 436 RUS 113/881/0
and is part of the European Integrated Infrastructure Initiative Hadron
Physics project under contract number RII3-CT-2004-506078.
A.~E.~is also supported by the Grants RFBR 06-02-16215 and 07-02-91557,
RF MSE RNP.2.2.2.2.6546 (MIREA) and by the Heisenberg-Landau Program of JINR.
The work was supported in part by DOE contract DE-AC05-06OR23177, under
which Jefferson Science Associates, LLC,  operates the Jefferson Lab.
F.~Y.~is grateful to RIKEN, Brookhaven National Laboratory and the U.S.\
Department of Energy (contract number DE-AC02-98CH10886) for providing
the facilities essential for the completion of this work.


\end{document}